\documentclass[twocolumn]{aastex701}

\usepackage{subcaption}

\begin{document}

\title{{Envelope Inflation and outflow Driven by Energy Deposition in Massive Stars}}

\shorttitle{}
\shortauthors{B. Mukhija and A. Kashi}

\author[0009-0007-1450-6490]{Bhawna Mukhija}
\affiliation{Department of Physics, Ariel University, Ariel, 4070000, Israel}
\email{bhawnam@ariel.ac.il}

\author[0000-0002-7840-0181]{Amit Kashi}
\email{kashi@ariel.ac.il}
\affiliation{Department of Physics, Ariel University, Ariel, 4070000, Israel}
\affiliation{Astrophysics, Geophysics, and Space Science (AGASS) Center, Ariel University, Ariel, 4070000, Israel}

\begin{abstract}
Evolved massive stars are known to undergo outflow with high mass ejections, resulting in the loss of a substantial portion of their envelopes. One proposed mechanism driving these events is the release or deposition of energy within the stellar envelope. We use a one-dimensional hydrodynamical code to investigate the resulting outflow and stellar response to energy deposition at specific regions inside a $\rm 70 \, M_{\odot}$ star. We compare hydrostatic and hydrodynamic models and test for different energies and widths of the depositing region.
We find that due to the deposited energy, the envelope expands significantly, and under certain conditions, such as assuming a uniform electron scattering opacity, this energy input becomes sufficient to unbind material from the outer envelope. This, in turn, leads to the formation of an  outflow.
We find that higher deposited energy triggers a strong outflow and results in a somewhat hotter and less expanded envelope due to the rapid loss of energy through expelled material.
This driving mechanism leads to sudden envelope expansion and the formation of strong outflows in our models, highlighting the generic hydrodynamic response of massive star envelopes to impulsive energy input.
\end{abstract}


\section{Introduction}

Evolved massive stars may experience outflow when energy is deposited into their envelopes, a process that has been proposed to play a key role in luminous blue variables (LBVs) eruptions \citep[e.g.,][]{1997ARA&A..35....1D, 2012ASSL..384.....D, 2016ApJ...817...66K, 2018Natur.561..498J,2023A&A...678A..55R, 2024ApJ...976..125D,2024ApJ...974..270C, 2024ApJ...974..124M,2025AJ....169..128S, 2025PASP..137j4201M, mukhija2025accretionrecoverygianteruptions, 2026NewA..12202475M,2025ApJ...994..159M, 2025Galax..13...29L}, and pre-supernova (pre-SNe) evolution of massive stars \citep[e.g.,][]{ 1997ARA&A..35..309F, 2004MNRAS.352.1213C, 2013Natur.494...65O, 2022ApJ...924...15J, 2024A&A...686A.231R}. \citet{2018MNRAS.480.1466S} argue that many aspects of the Giant Eruptions (GEs) and the structure of the resulting nebula cannot be fully explained by a steady, super-Eddington wind alone. Instead, they suggest that a more explosive component resembling a SNe-like event is required in addition to the outflow\footnote{In this context an “outflow” denotes a transient, eruptive mass ejection events, and should not be confused with the steady line-driven winds characteristic of massive stars during their quiescent evolution.}. Similar to what has been proposed for pre-SN eruptions (see the review by \citet{2014ARA&A..52..487S} and references therein), this explosive outflow phase likely interacts with an extended period of quasi-steady wind-driven outflow. Instabilities in the stellar core may trigger such pre-SN eruptions during the final stages before collapse. Later on, it can deposit energy into the envelope through mechanisms such as gravity waves \citep[e.g.,][]{2011ApJ...738L...5P, 2012MNRAS.423L..92Q, 2018MNRAS.476.1853F}, or directly into the core through processes like pair-instability pulsations \citep[e.g.,][]{1967PhRvL..18..379B, 2007Natur.450..390W, 2011ApJ...734..102K, 2016MNRAS.456.1320T, 2017ApJ...836..244W}, or by late shell-burning instabilities \citep{2014ApJ...785...82S}.

One possible mechanism for these outflows is super-Eddington energy deposition.  Such energy deposition can occur at specific radii within the star and be triggered by several distinct processes. These processes are: tidal heating or via a companion in binary systems during the common envelope evolution (CEE) by \citet[e.g.,][]{1976IAUS...73...75P,2018MNRAS.475.1198S} or by non-local redistribution of energy i.e., wave-driven energy transport in the core due to unstable fusion by \citet[e.g.,][]{2011ApJ...738L...5P,2012MNRAS.423L..92Q}. 
However, the additional energy sources responsible for triggering these events, as well as the dynamical mechanisms driving outflows and their observational signatures, remain not fully understood.

Several studies have explored the dynamics of massive star envelopes in response to external energy deposition \citep[e.g.,][]{2016MNRAS.458.1214Q, 2017MNRAS.470.1642F, 2018MNRAS.476.1853F, 2019ApJ...877...92O, 2019MNRAS.485..988O, 2020A&A...635A.127K}. \citet{2016MNRAS.458.1214Q} investigate how a star responds to continuous energy deposition near its surface at rates exceeding the Eddington luminosity, the point at which radiation pressure overcomes gravity. Their focus is on the star's response over a thermal timescale, leading to the formation of an extended convective zone in the outer layers. This zone, where energy is transported by convection rather than radiation, plays a key role in driving the outflow. They explore the quasi-hydrostatic response of the star, where its outer layers expand and adjust without undergoing dynamical collapse. Their work highlights how convective energy transport helps regulate the star's structure, driving outflows that are stronger than typical radiation-driven winds, especially in stars with extreme luminosities, such as LBVs and Wolf-Rayet (WR) stars. Another study by \citet{2019ApJ...877...92O} explored sudden energy deposition in the stellar envelope, typically focusing on cases where the deposited energy is comparable to the stellar binding energy.
They found that highly super-Eddington energy input leads to unrealistic SN properties, while sub-Eddington injection causes moderate inflation, which may trigger outflow via secondary effects like pulsations or binary interactions. \citet{2020A&A...635A.127K, 2021A&A...646A.118K} explored the nature of eruptive mass-loss events in massive stars, via connection to energy injection at the base of the stellar envelope, originating from nuclear burning in the core. Rather than detailing the specific physical mechanisms responsible for transporting this energy to the envelope, their study adopts a parametric approach, characterizing the amount and duration of energy injection and comparing the outcomes with observational data. They examine the pre-SNe evolution of various stellar progenitors, including red, yellow, and blue supergiants as well as WR stars. Their study presents predicted light curves associated with eruptive episodes, estimates of the mass ejected, and the structure of the circumstellar material (CSM) at the time of core collapse (CC). Their findings suggest that a rapid injection of energy at the base of the envelope on a timescale shorter than the envelope’s dynamical timescale can replicate optical outflow observed before CC and generate CSM capable of powering a Type IIn SNe through the ejecta-CSM interaction.


In \citet{2024ApJ...974..124M}, we also studied giant eruptions events in $\rm 70 \, M_{\odot}$ star during the post-MS phase. However, our model was based on calculations where outflow event was manually implemented in \textsc{mesa}. We removed mass from the outer layers of the star, assuming that the event was triggered by energy deposition from a binary companion. Our calculations do not involve energy deposition to induce the outflow event. As a result of the outflow, the luminosity decreases by an order of magnitude, and the star contracts, shifting towards the hotter side of the HR diagram. Recently, \citet{2024ApJ...967...33C} used 1D hydrodynamical simulations, and investigated how sudden energy deposition in the stellar interior influences the outflow, analyzing its dependence on both the amount and location of the deposited energy. They found that the ejected mass exhibits a complex relationship with deposition energy, driven by the breakout of density and pressure waves. These waves, generated by the energy injection, propagate through the star and interact via reflections from the core, reaching the surface, and driving the outflow. However, their model does not include the energy transport by radiation diffusion.

\begin{figure}
    \centering
    \includegraphics[width=\linewidth]{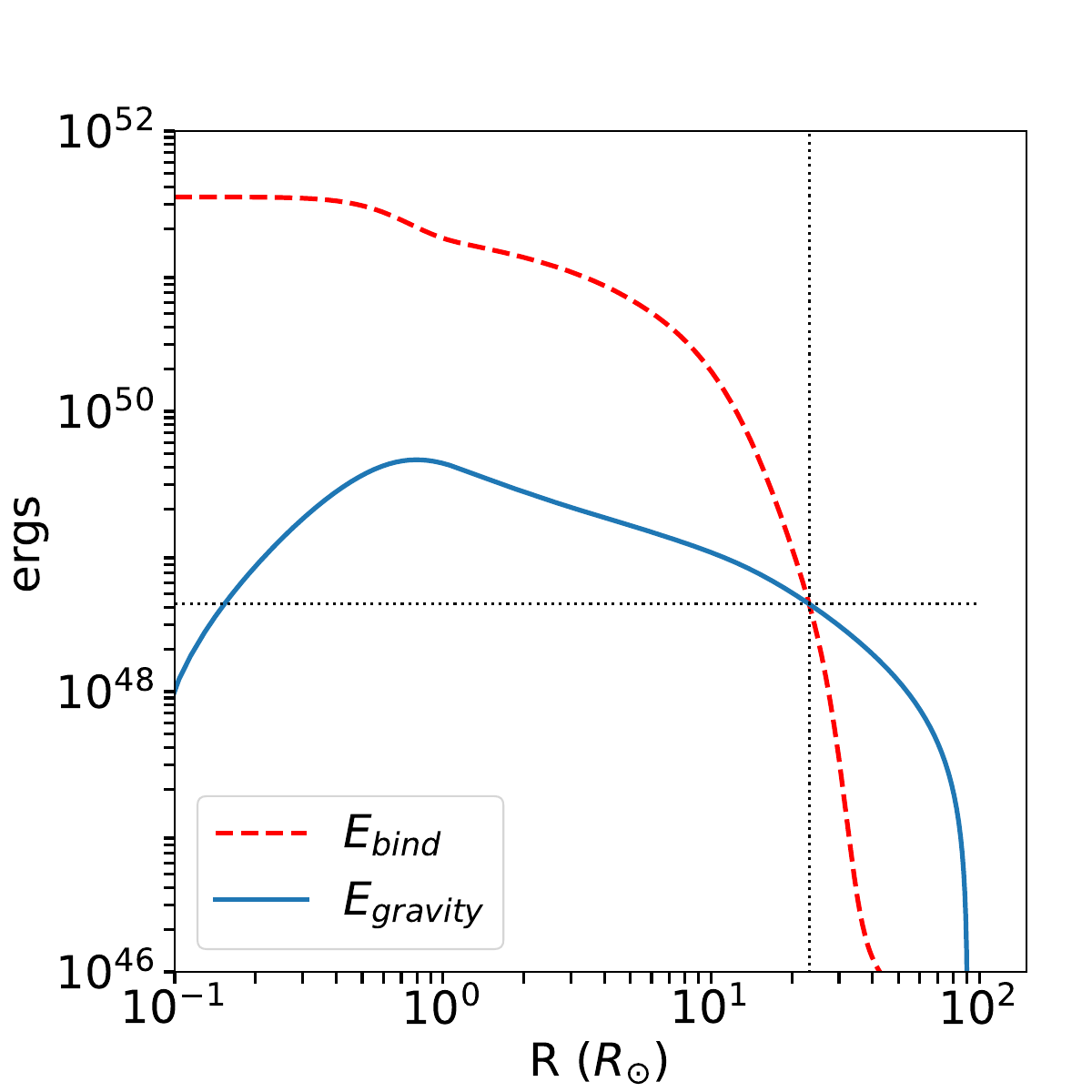}
    \caption{The binding energy and gravitational energy of our model are shown as functions of the stellar radius ($R$). In this setup, the primary star has a ZAMS mass of 70 $\rm M_{\odot}$, while the companion is treated as a point mass. The point of intersection, where both energies become equal, occurs at a critical radius of $R_{\rm crit} = $ 23.21 $\rm R_{\odot}$, corresponding to an energy of $4.22 \times 10^{48}$ erg.}
    \label{fig_1}
\end{figure}

In this work, we investigate an idealized scenario in which energy is deposited instantaneously within a narrow region of the stellar interior. This setup assumes that the timescale for energy injection is shorter than the local dynamical timescale\footnote{Our setup assumes that the timescale for energy injection is shorter than the local dynamical timescale, corresponding to an impulsive (“thermal-bomb”) energy deposition in which the full injected energy is added instantaneously and the subsequent envelope response is followed. We ran our simulations with time steps much shorter than the stellar dynamical timescale to accurately capture the dynamical response of the envelope.} Our goal is to quantify the total mass lost from the star as a function of the amount of deposited energy $E_{\rm dep}$, and the width of the heating regime. By employing 1D hydrodynamic simulations, we aim to establish a fundamental understanding, or a `base case' of how the stellar envelope responds to energy input. Later on,  we investigate the response of a star to energy deposition near its surface by examining its impact on the stellar structure. In real systems, such energy deposition may arise from processes such as wave-driven energy transport from the core, binary interactions, or accretion-powered feedback during outflow phases. Observationally, LBV eruptions and pre-supernova outbursts exhibit characteristic timescales ranging from months to years, with ejected masses spanning 10--20 $\rm M_{\odot}$ and  mass-loss rates up to $\sim 1 ~\rm M_{\odot}\, yr^{-1}$, as exemplified by extreme events such as $\eta$~Carina \citep[e.g.,][]{1999PASP..111.1124H, 2012Natur.486E...1D, 2012ASSL..384....1H, 2006MNRAS.372.1133G}. In our calculations, we adopt an energy-injection duration of 1.2 years, motivated by the fact that our stellar models are already highly extended, with radii of order $10^{4}\, \rm R_{\odot}$, for which the dynamical timescale is correspondingly long. We do not aim to reproduce any specific observed event. Instead, this deliberately simplified setup is designed to probe limiting cases of envelope response and to isolate the fundamental hydrodynamic behavior of the stellar envelope under localized energy deposition, independent of the detailed physical triggering mechanism.

\citet{2024ApJ...974..270C} presented a physically motivated mechanism for modeling eruptive outflow in massive stars, based on the criterion that a local super-Eddington luminosity can unbind overlying stellar layers when the radiative energy surpasses their gravitational binding energy. Their implementation in \textsc{mesa} introduces a mass-loss scheme that activates in regions of the envelope with optical depth below a critical value. This approach results in substantial mass-loss rates, up to $\sim 10^{-2}~\rm M_\odot\,{\rm yr}^{-1}$, particularly during the post-MS evolution. One of the key outcomes of their work is the suppression of Red Supergiant (RSG) formation for stars above $\sim \rm 20~M_\odot$, providing a potential explanation for the observed lack of high-mass RSGs. In comparison, we study the stellar envelope's response to localized energy deposition within the stellar interior using the \textsc{mesa} code. Rather than relying on internal luminosity conditions, our method involves the artificial injection of energy over prescribed spatial and temporal domains, motivated by physical scenarios such as binary interaction or instabilities. Our models examine both hydrostatic and hydrodynamic cases and assess how different deposition depths and rates influence envelope expansion and mass outflow. Although both studies investigate eruptive outflow in massive stars using 1D stellar evolution models, they differ in motivation and methodology. \citet{2024ApJ...974..270C} develop a luminosity-driven, self-regulating prescription linked to super-Eddington regions, while we focus on the consequences of sudden, externally energy input. These findings complement each other by highlighting both steady and impulsive pathways for enhanced outflow in evolved massive stars.

The paper is organised as follows. The basic assumptions and method of modeling are discussed in section \ref{2}.  The simulations and results are described in section \ref{3}. Our discussion and summary are given in section \ref{4} and \ref{5} respectively.

\section{Model setup}
\label{2}

\subsection{Physical ingredients for the modeling}
\label{2.1}

We use the 1D hydrodynamical stellar evolution code \textsc{mesa} \citep[e.g.,][]{2011ApJS..192....3P, 2013ApJS..208....4P, 2015ApJS..220...15P, 2018ApJS..234...34P, 2019ApJS..243...10P}  
 to study the star's response to energy deposition in its interior to the surface.
Our approach does not aim to produce self-consistent stellar evolution models, but instead focuses on highlighting the numerical response of the envelope as the treatment transitions from hydrostatic to a hydrodynamical regime after the deposition of the energy. In \textsc{mesa}, an artificial energy deposition that produces an outflow or shock wave can be triggered using one of three methods: a piston, a luminosity flash, or a thermal bomb \citep{2018ApJS..234...34P}. The piston method involves altering the inner boundary conditions, specifically the velocity and radius, to simulate the movement of a piston. This approach is often used to mimic a CCSN, where the inner boundary first moves inward at free-fall velocity, then rapidly reverses direction to simulate the rebound and explosion. The piston driven explosion is parameterized following the formalism of \citet{1995ApJS..101..181W}, using key parameters such as the duration of the infall phase, the minimum radius reached by the piston, the initial velocity of the outward motion, and the final radius of the piston after the explosion.
The second method, known as the luminosity flash, involves injecting energy by temporarily increasing the luminosity at the inner boundary over a defined time interval to achieve a targeted total energy input. During this process, the inner boundary radius is held constant. Once the luminosity flash concludes, the boundary condition switches to a zero-flux state. The injected energy is deposited into the first computational zone adjacent to the inner boundary \citep{2018ApJS..234...34P}.
The third approach, the thermal bomb, deposits energy at a constant rate over a set time interval within a specific region of the star, defined by two chosen Lagrangian mass coordinates. According to \citet{2018ApJS..234...34P}, the selected method for initiating the outflows can substantially influence the resulting characteristics of the stellar envelope.
They also outlined that the behavior of the outflow and the resulting explosive nucleosynthesis depend on both the amount of energy deposited and the timescale of deposition. However, the overall shock dynamics are primarily governed by the total deposited energy.

Accurately modeling stellar outflows in hydrodynamic simulations requires extremely small timesteps. To prevent unphysical large accelerations, it is necessary to limit the growth of convective velocities predicted by the standard instantaneous mixing length theory. As highlighted by \citet{2018ApJS..234...34P}, allowing convective velocities to respond instantaneously can cause outflow models to fail, as even rapid energy deposition in a localized region would be quickly transported away by convection, suppressing the formation of outflows.
Therefore, to effectively simulate outflows, it is crucial to impose a restriction on how quickly convective velocities can accelerate. Thus, we use the time-dependent convection theory (\textsc{tdc}) to handle the convection during the outflow with mixing length parameter $\alpha_{\rm MLT}$ = 1.6 \citep{2023ApJS..265...15J}. When simulating the interior of the star, while maintaining low surface velocities, one common approach is to adopt a surface pressure based on a selected atmospheric model. This method uses the momentum equation to link the surface velocity with the gradient of surface pressure. \textsc{mesa} adopts a diffusive approach to model semiconvection, using a diffusion coefficient based on the method outlined by \citet{1983A&A...126..207L}. In our models, we apply a fixed semiconvection efficiency factor of 1, consistent with previous studies \citep[e.g.,][]{2006A&A...460..199Y, 2019A&A...625A.132S}. Core convective overshooting is also included, implemented with an exponential decay profile and an overshooting parameter of $ f(\rm 1) = 0.01$ and $ f_{0}(\rm 1) = 0.001$ following the prescription by \citet{2000A&A...360..952H}.

 \begin{table}
    \centering
    \begin{tabular}{l c}
    \hline 
    \hline
     $M_{ \rm ZAMS} ~( \rm M_{\odot})$ & 70 \\
     \hline
     $ \rm star~age~ (Myrs) $ & 2.918736 \\
     $M ~( \rm M_{\odot})$ & 70   \\ 
     $\log T_{\rm eff} ~(\rm K) $& 4.299   \\
      $\log R ~(\rm R_{\odot}) $ & 1.956 \\
     
     $\log \rho ~(\rm g~cm^{-3})$ & -0.102  \\
     $\log L ~(\rm L_{\odot})$ & 5.066\\

   \hline \hline
    \end{tabular}
    \caption{Stellar parameters correspond to the initial stage of our star when the experiment starts. The rows give the star age, the mass of the star ($ M$), surface temperature ($ T_{\rm eff}$), density ($ \rho$), and surface luminosity ($ L$) respectively.}
    \label{T1}
\end{table}

To study the stellar response to $E_{\rm dep}$, we first evolve the $\rm 70 \, M_{\odot}$ star to the post-main sequence (post-MS) phase without applying any additional energy input. The stellar parameters at this stage are listed in Table \ref{T1}. At this point, we introduce energy deposition.
The starting point of the energy deposition on the evolutionary track after the post-MS is chosen as the point where $T \simeq 19\,400 $ K, which corresponds to the place on the LBV instability strip for $L\simeq 10^6 {L_\odot}$  \citep{2017RSPTA.37560268S}. The choice to identify this point for the start of $E_{\rm dep}$ is backed by the observed placement of LBVs like $ \eta$- Carina  \citep[e.g.,][]{1999PASP..111.1124H, 2006AJ....132.2717M, 2007ApJ...666.1116S,2008MNRAS.390.1751K, 2010MNRAS.405.1924K, 2010ApJ...717L..22M, 2012Natur.486E...1D, 2012ApJ...751...73M, 2012ASSL..384....1H, 2014A&A...564A..14M} and P Cygni \citep[e.g.,][]{1988IrAJ...18..163D,1994PASP..106.1025H, 1997A&A...326.1117N,1999SSRv...90..493I, 1999ASPC..192...32D,2001ASPC..233..133N, 2010AJ....139.2269B, 2010MNRAS.405.1924K, 2012ASSL..384...43D,2018NewA...65...29M, 2020MNRAS.494..218R} in this region on the HR diagram, where they undergo the eruption phase and sheds a significant mass from their envelope. We introduce energy deposition under two distinct cases, with $E_{\rm dep}=4.22\times10^{48},\rm erg$ corresponding to the intersection point of the binding and gravitational energies at a  critical radius ($R_{\rm crit}$) of 23.21 $\rm R_\odot$,  as shown in Fig~\ref{fig_1}.

In Case 1 (section \ref{3.1}), we include the $\textsc{mlt++}$ treatment, which suppresses outflow generation, and we neglect electron scattering. This setup represents the hydrostatic response of the stellar envelope to the energy deposition. In Case 2A (section \ref{3.2}), we disable the $\textsc{mlt++}$ treatment and artificially enhance electron scattering by applying a scattering factor of 100, not as a representation of a physical opacity source, but as a numerical means to enforce a radiation-pressure–dominated envelope and to examine the hydrodynamical response under extreme conditions. This configuration allows us to model the hydrodynamical response of the star to the deposited energy. The use of \textsc{mlt++} in hydrostatic calculations artificially suppresses outflow generation, as it enhances convective energy transport, thereby preventing the buildup of energy required to drive outflows. For our hydrodynamical calculations, we evolve the star without \textsc{mlt++} during the energy deposition phase. Additionally, in this framework, outflow is not imposed through a prescribed wind recipe; instead, it is inferred from dynamically ejected material that becomes energetically unbound following the energy injection. In Case~1 and Case~2A, the values of \(E_{\mathrm{dep}}\) and the width of the heating region are the same as those listed in Table~\ref{T2}. Later, we extended our study by introducing Cases 2B (section \ref{3.3}) and 2C (section \ref{3.4}), in which we varied the width of the heating region and the value of the deposited energy, respectively, relative to Case 2A.

\subsection{Calculations for the energy deposition value and its deposition region}
\label{2.2}

We select the region for energy deposition based on calculations involving the binding energy and gravitational energy of the stellar envelope. In this section, we outline the energy considerations relevant to the process for envelope's mass ejection. We consider a binary interaction only as a scaling argument to estimate the amount of energy that may be deposited into the envelope of the primary star. In particular, we compare the injected energy to the orbital gravitational energy that would be released if a point-mass companion were to spiral inward to a separation at which this energy becomes comparable to the envelope binding energy. However, in our calculations, we model a single star, and the binary interaction itself is not followed dynamically. The energy deposition is therefore parameterized and may represent tidal heating associated with a companion, rather than a self-consistent binary evolution. Following \citet{2011MNRAS.417.1466K}, we assume that once the envelope material interior to the companion reaches co-rotation with the orbital motion, the inspiral continues. This is because the envelope material exterior to the companion cannot synchronize, leading to strong tidal interactions. Based on this scenario, we assume that an amount of energy roughly equal to the envelope’s binding energy is deposited over a short timescale. Accordingly, we present the expressions for both the orbital gravitational energy and the envelope binding energy below. The mass inward to radius $r$ is given by

\begin{equation}
    M( r) \approx M_{\rm core} + M_{\rm env}(r)
\end{equation}

\begin{equation}
   M(r) \approx M_{\rm core} + \int_{\rm R_{\rm core = 0}}^{r} 4 \pi r^{2} \rho dr
\end{equation}

Our post-MS stellar mass is $M$ = $\rm 70 ~M_{\odot}$, its core mass is $M_{\rm core}$ = $\rm 27.56 ~M_{\odot}$, its envelope mass is $M_{\rm env}$ = $\rm 42.44 ~M_{\odot}$ and the stellar radius is $R$ = $\rm 90.36~R_{\odot}$. Thus the binding energy of the envelope mass outside radius $r$ is given by

\begin{equation}
    E_{\rm bind} =\frac{ \int_{r}^{R_{*}} G [M(r) + M_{2}(r)]}{r} 4\pi r^{2} \rho dr
\end{equation}

Here we have taken into account the companion mass ($M_{\rm 2}$) as a point mass, and neglected the thermal energy of the gas. 
This choice is motivated by two considerations. First, adopting a more massive companion would increase the available orbital energy and thus the amount of energy required to unbind the envelope in our scaling estimates. In practice, such larger energy injections can lead to numerical convergence difficulties in \textsc{mesa}. For this reason, we adopt a low-mass point companion for computational robustness and convenience. We nevertheless explore higher energy inputs by directly increasing the deposited energy in dedicated runs (e.g., Case 2C), thereby capturing the dynamical response expected for more energetic interactions without explicitly increasing $M_{\rm 2}$. Second, introducing a higher-mass companion would require accounting for its internal structure and thermal energy, as well as for additional interaction physics, which would substantially increase the complexity of the system and move beyond the scope of our idealized framework.

We follow the \citet{2011MNRAS.417.1466K}, and assume that as the companion enters the envelope, its primary influence is through tidal interactions that spin up the envelope. This spin-up, along with the companion's penetration into the envelope, is likely to induce some outflow. The gravitational energy released by the companion as it spirals inward from the stellar surface to a final orbital separation $a$ is given by

\begin{equation}
    E_{\rm G} = \frac{G M_{a} M_{2}}{2a} - \frac{G M_{*} M_{2}}{2 R_{*}}
\end{equation}
\begin{figure*}[htbp]
    \centering
    \begin{minipage}[t]{0.33\textwidth}
        \centering
        \includegraphics[width=1\linewidth]{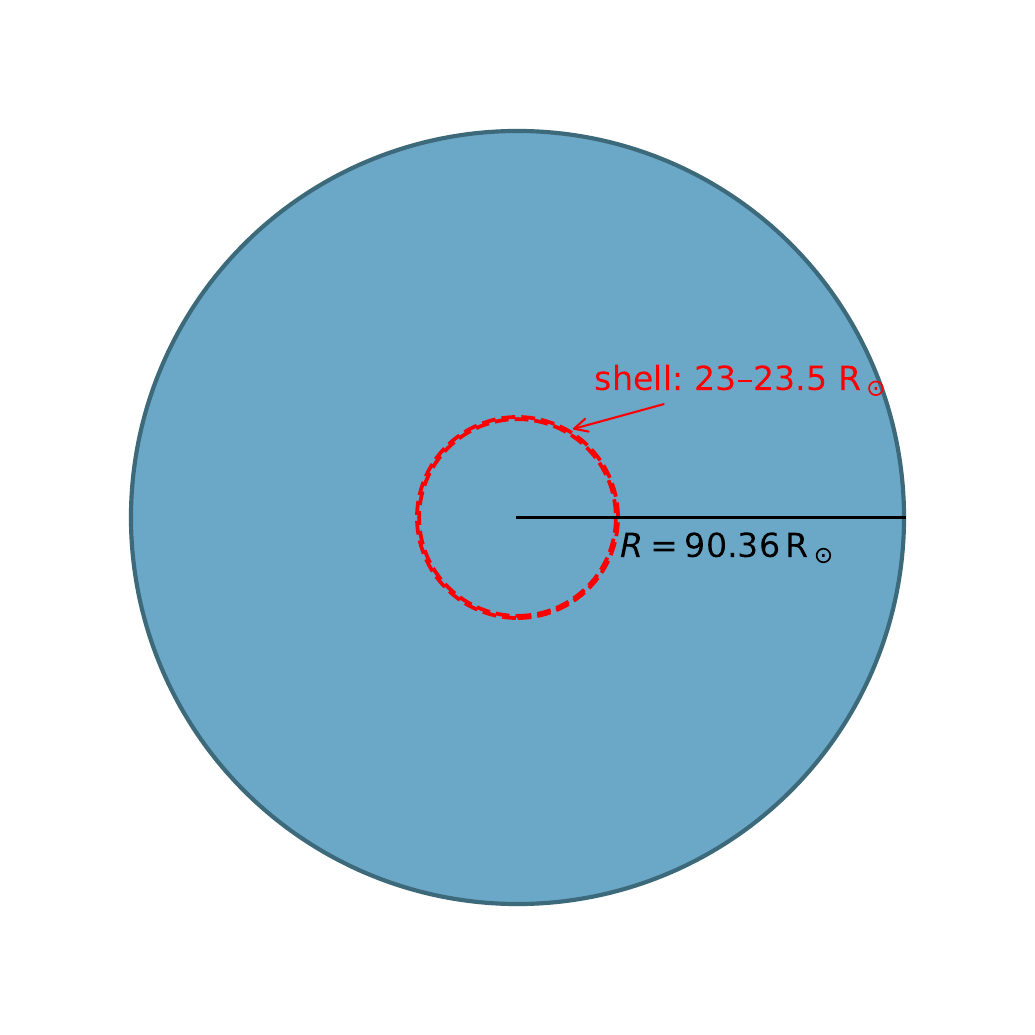}
        \textbf{(a)}
    \end{minipage}\hfill
    \begin{minipage}[t]{0.33\textwidth}
        \centering
        \includegraphics[width=1\linewidth]{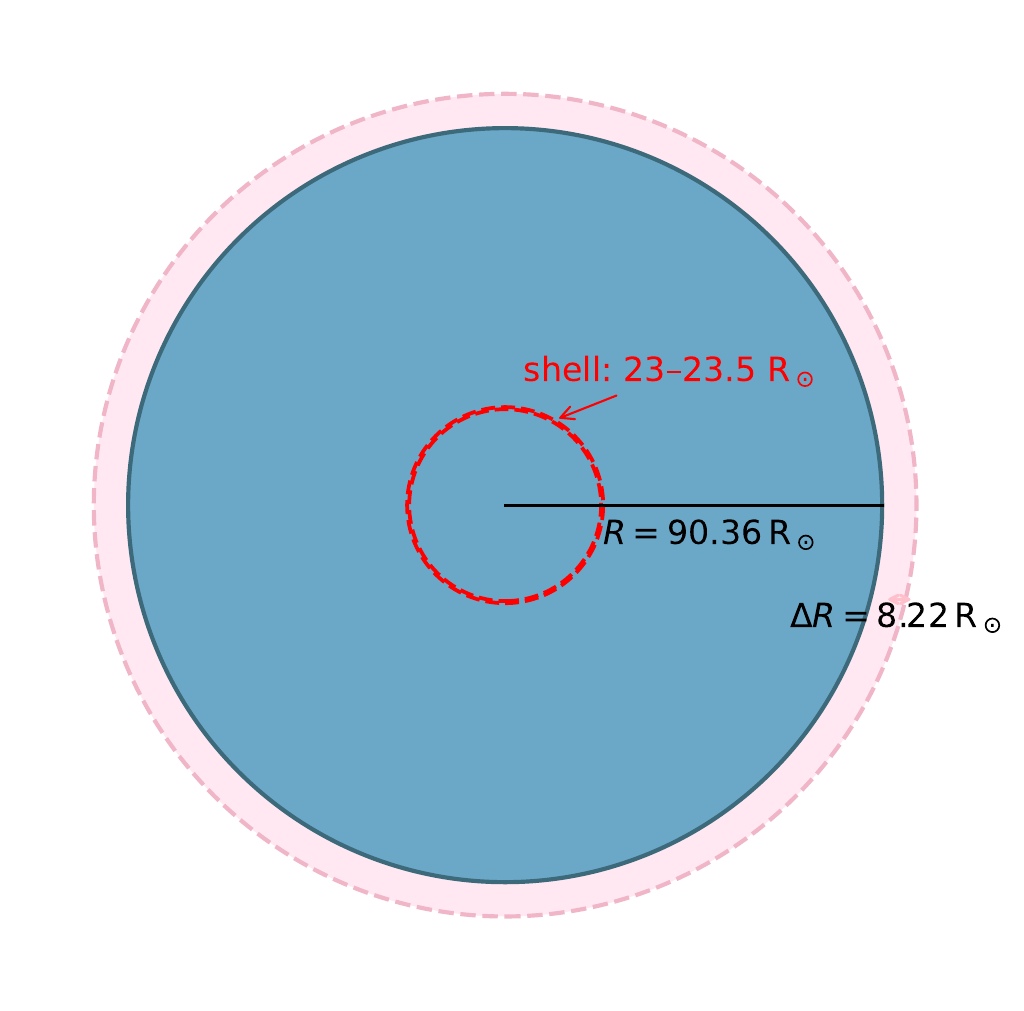}
        \textbf{(b)}
    \end{minipage}\hfill
    \begin{minipage}[t]{0.33\textwidth}
        \centering
        \includegraphics[width=1\linewidth]{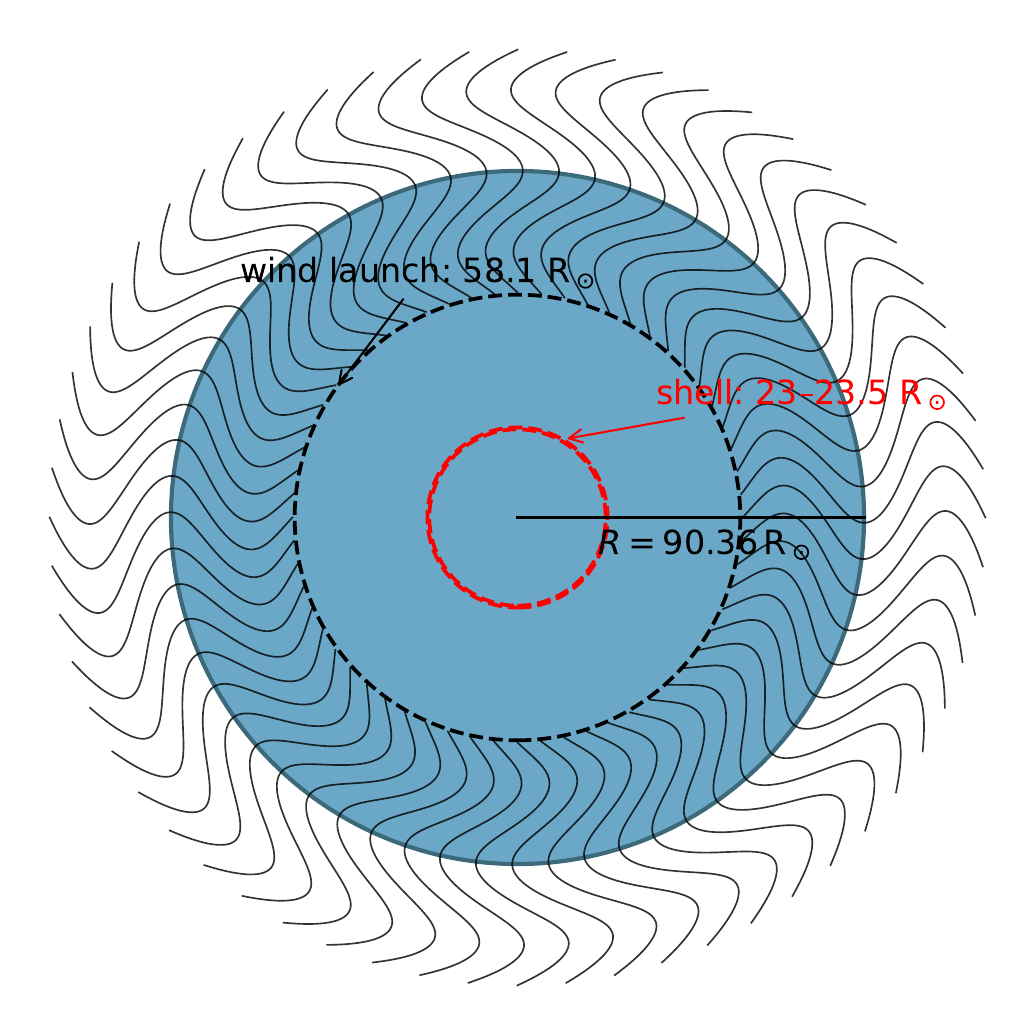}
        \textbf{(c)}
    \end{minipage}

    \caption{Schematic diagram of a $70\,\rm M_{\odot}$ star. 
    \textbf{(a)} Stable pre-deposition stage; the stellar parameters are listed in Table~\ref{T1}. 
    \textbf{(b)} Hydrostatic response (Case~1): the envelope expands after energy deposition, with the deposition shell indicated. 
    \textbf{(c)} Hydrodynamic response: outflows are launched near radius $R_{\rm out} \simeq \rm 58.1\,\rm R_{\odot}$ and propagate outward to $\sim 10^{4}\,\rm R_{\odot}$.}
    \label{fig2}
\end{figure*}

The binding energy and the gravitational energy released by the companion for our model are shown in Figure \ref{fig_1}, assuming $M_{2}$ = $1\rm ~M_{\odot}$. At the critical radius of $R_{\rm crit}$ = $23.21~\rm R_{\odot}$, the two energies become equal. To trigger outflow. Based on these calculations, we adopt a deposited energy, as mentioned in Table~\ref{T2} for all cases, and time intervals, and assume that a heating source is present within the region between 23 to 23.5 $ \rm R_{\odot}$.
In \textsc{mesa}, this energy is introduced as heat using the \texttt{other\_energy} hook within the \texttt{run\_star\_extras} module. The width of the energy deposition region is set to $\rm 0.5~R_{\odot}$, a value chosen based on our modeling assumptions. Here, the choice of heating region width, following \citet{2018ApJS..234...34P} (section 4.9), where the energy is deposited over a narrow Lagrangian mass range of 0.5 $\rm M_{\odot}$, we adopt the same assumption in our calculations. In our stellar model, this corresponds to a radial extent of about 0.5 $\rm R_{\odot}$, which we therefore use as the width of the heating region. After depositing the energy, we analyze the stellar response for a duration of 1.2 years with an interval of $t = 0.4$ years. We stress that the adopted energy-deposition prescriptions do not constitute a physically self-consistent model of the outflow mechanisms. Instead, they represent controlled, artificial perturbations designed to explore limiting hydrodynamic responses of the envelope. Consequently, the resulting outflows should be interpreted as idealized envelope ejections, and not as self-regulated outflow, and thresholds therefore reflect the assumptions of this simplified framework and may differ from those of real stellar systems.

The adopted duration of 1.2 years for the energy deposition is not intended to represent the physical timescale of LBV eruptions, which typically span several years. Instead, it is dictated by numerical and structural constraints of our stellar models: once the envelope expands to radii of order $10^4~\rm R_{\odot}$ the \textsc{mesa} models become increasingly unstable and cannot be evolved self-consistently for much longer timescales. The chosen duration therefore, reflects the maximum interval over which the hydrodynamic response can be followed in a controlled manner within our framework. The intermediate sampling interval of 0.4 years is similarly not meant to correspond to any specific physical cadence. It is adopted purely for diagnostic purposes, providing sufficient temporal resolution to track the evolution of the outer-layer profiles while maintaining clarity and consistency.

\begin{table}[t]
    \centering
    \begin{tabular}{llcccc}
        \toprule

        Case & HS/ & $E_{\rm dep}$ & $R_{\rm dep}$ & $\delta R$ & Sec.  \\
        \ & HD & ($10^{48}$ erg) & ($R_{\odot}$)  & ($R_{\odot}$)   \\
        \hline
        Case 1 & HS & 1.45 & 23-23.5 &0.5 & \ref{3.1} \\
        Case 2A & HD & - & &0.5 & \ref{3.2}\\
         t (yrs) &0.4 & 0.48  & 23-23.5\\
         & 0.8& 0.96 & 23-23.5\\
          & 1.2& 1.45 & 23-23.5\\
        Case 2B & HD & - & & 0.8 & \ref{3.3} \\
         t (yrs) &0.4& 0.48 & 23-23.8\\
         &0.8&0.96 & 23-23.8\\
          & 1.2&1.45 & 23-23.8\\
        Case 2C & HD & - & &0.5 & \ref{3.4}\\  
         t (yrs) &0.4& 0.72 & 23-23.5\\
         & 0.8& 1.44 & 23-23.5\\
          & 1.2&2.17 & 23-23.5\\
     \hline
    \end{tabular}
       \caption{ Input parameters for different cases, including the energy deposited in the star ($E_{\rm dep}$), heating region ($R_{\rm dep}$), and width of the heating region ($\delta R$) for the hydrostatic (HS) and hydrodynamic (HD) case.}
    \label{T2}
\end{table}

\begin{figure*}
    \centering

    \begin{subfigure}{\textwidth}
        \centering
        \includegraphics[trim={0.8cm 0cm -0.0cm 0.0cm},clip,width=1\textwidth]{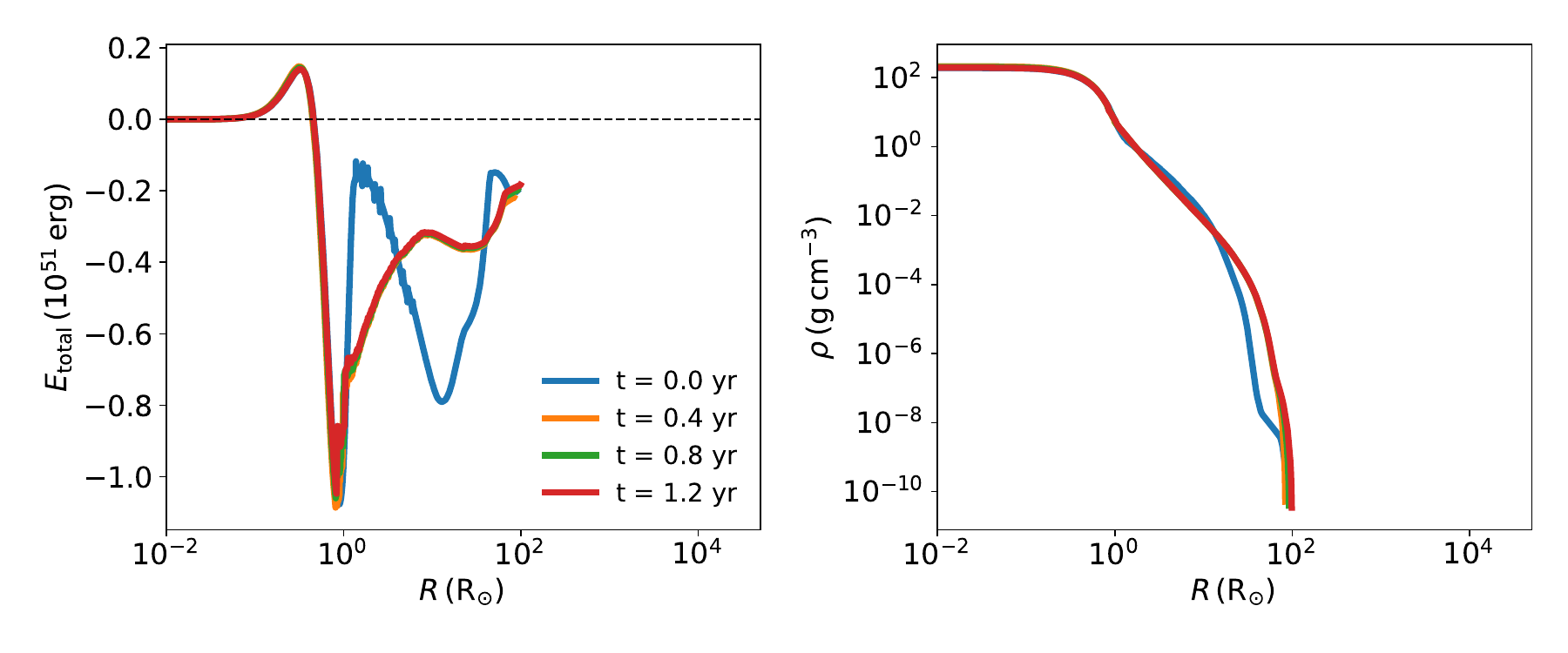}
        \caption{ Case 1: showing the hydrostatic response of the stellar envelope to energy deposition. It represents the evolution of a physical quantity: total energy, and density as a function of radius at several time snapshots. The figure illustrates how the star adjusts by expanding its envelope without developing an outflow, consistent with hydrostatic adjustment.}
       
    \end{subfigure}

    \vspace{0.5cm}

    \begin{subfigure}{\textwidth}
        \centering
        \includegraphics[trim={0.0cm 0.0cm -0.0cm 0.0cm},clip,width=1\textwidth]{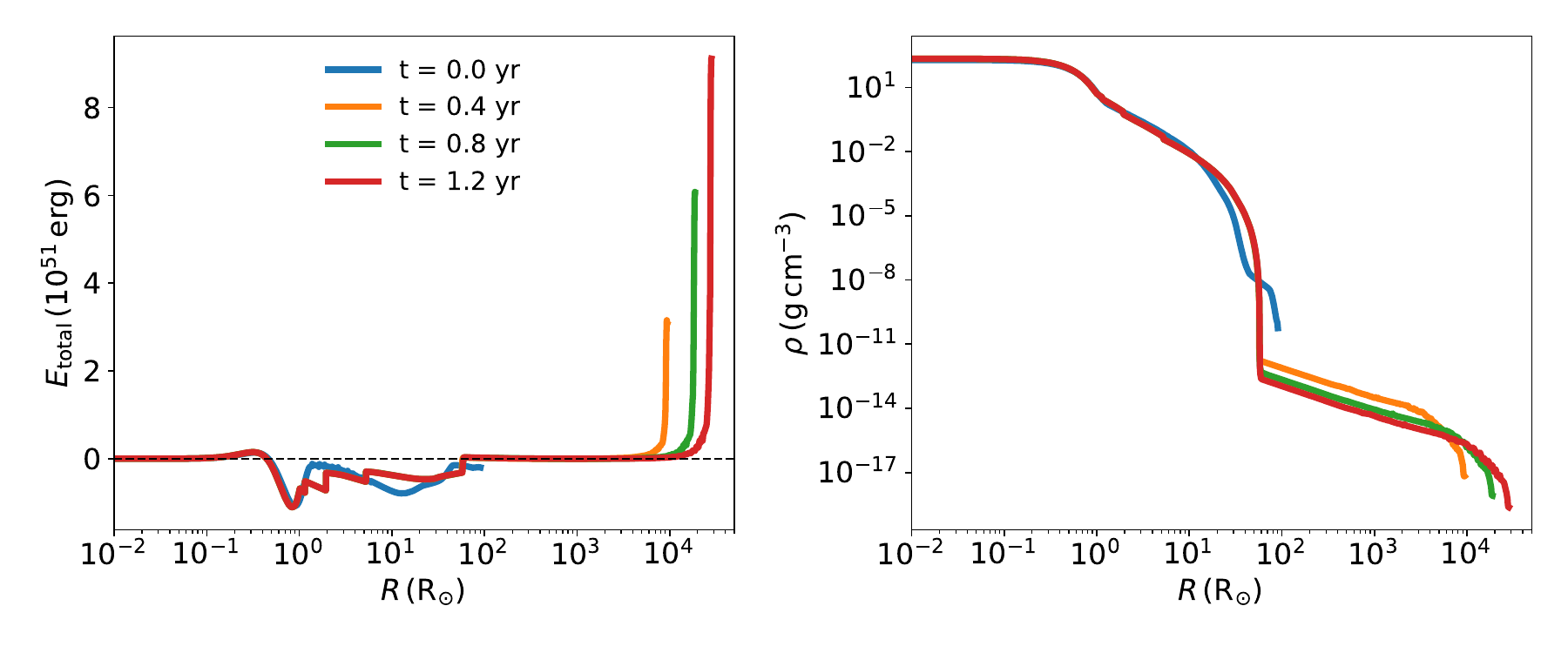}
        \caption{Case 2A: showing the hydrodynamic response of the stellar envelope to energy deposition. It represents the evolution of a physical quantity: $E_{\rm k}$ of the outflow, and density as a function of radius at several time snapshots. The figure illustrates how the star adjusts by expanding its envelope with the development of an outflow.}
        
    \end{subfigure}

    \caption{
   Panel (a) shows the hydrostatic, while panel (b) shows hydrodynamic response of the stellar envelope to energy deposition.
    }
    \label{fig3}
\end{figure*}

\section{Results}
\label{3}

\subsection{Case 1: Hydrostatic adjustment in response to the energy deposition}
\label{3.1}

In this case, the heating source is present within the region between 23 to 23.5 $ \rm R_{\odot}$, with an energy deposition value mentioned in Table~\ref{T2}.
As a result of the energy deposition, the stellar envelope begins to expand outward due to the increased thermal pressure. The values of energy deposition which induces convection, and works to transport the energy that photons cannot carry \citep{1973ApJ...181..429J}. In this hydrostatic model (i.e., without outflows), this results in the formation of a large envelope, even if the star was initially compact. Initially, at $t=0$ years, the stellar radius is 90.36 $\rm R_{\odot}$. Following the energy deposition, the stellar radius expands to $ R_{\rm ext}$ = 98.58 $\rm R_{\odot}$ after 1.2 years. This represents an increase in radius by a factor of about 1.089. Here, luminosity decreases by a factor of 0.71, and temperature decreases by a factor of 0.78 due to the expansion. 
Since the outflow is initiated as a thermal bomb, most of the energy is initially stored as internal energy. As the material expands and accelerates, kinetic energy ($E_{\rm k}$) increases while internal energy correspondingly decreases. This implies that, during this phase, no energy is utilized to unbind the envelope. As shown in Figure \ref{fig3} panel (a) during the hydrostatic case, the total energy in the outer layers of the star remains negative, confirming that the material in these regions is still gravitationally bound even after the energy deposition; there is no outflow in this scenario. While density within the extended envelope reaches down to $10^{-10}~ \rm g \,cm^{-3}$. However, the change is minimal even after the deposition, indicating that the effect is negligible in the outer layer. This suggests that no outflow is launched in this scenario, only the envelope expanded by the $\Delta R = 8.22 \, \rm R_{\odot}$ as the result of energy deposition. The schematic diagram of this case is shown in Figure~\ref{2} panel (b), while panel (a) shows the stable 70 $\rm M_{\odot}$ star.


\subsection{Case 2A: Time-Dependent Hydrodynamics with \textsc{mesa}}
\label{3.2}

In this case, we use the implicit hydrodynamics capabilities of the \textsc{mesa} to compute a solution for a star's hydrodynamic response to energy input. Unlike the calculations in earlier work, which are hydrostatic and could not produce a outflow, this analysis incorporates hydrodynamic effects. The key aspects of this model include the use of \textsc{mesa} implicit hydrodynamics capabilities, which involve solving the time dependent, spherically symmetric hydrodynamic equations. These calculations incorporate radiation in the diffusion approximation under the assumption of local thermal equilibrium. Unlike in section \ref{3.1} (Case 1), we do not use the \textsc{mlt++} convection module, as it would overestimate the efficiency of convective energy transport. Additionally, we disable \textsc{mesa} wind recipes to ensure mass conservation within the domain. A key simplification in our approach is the assumption of a uniform electron scattering opacity throughout the model with a value of 100. Due to the limitation on convective accelerations, the deposited energy increases the internal energy within the innermost layer of our grid. This buildup of pressure triggers a rapid expansion of the inner layers, leading to the formation of the outflows.

\begin{figure*}[t]
  \centering
  \begin{tabular}{c @{\qquad} c }
    \includegraphics[trim={0.5cm 0.0cm -0.3cm 0.2cm},clip,width=0.5\textwidth]{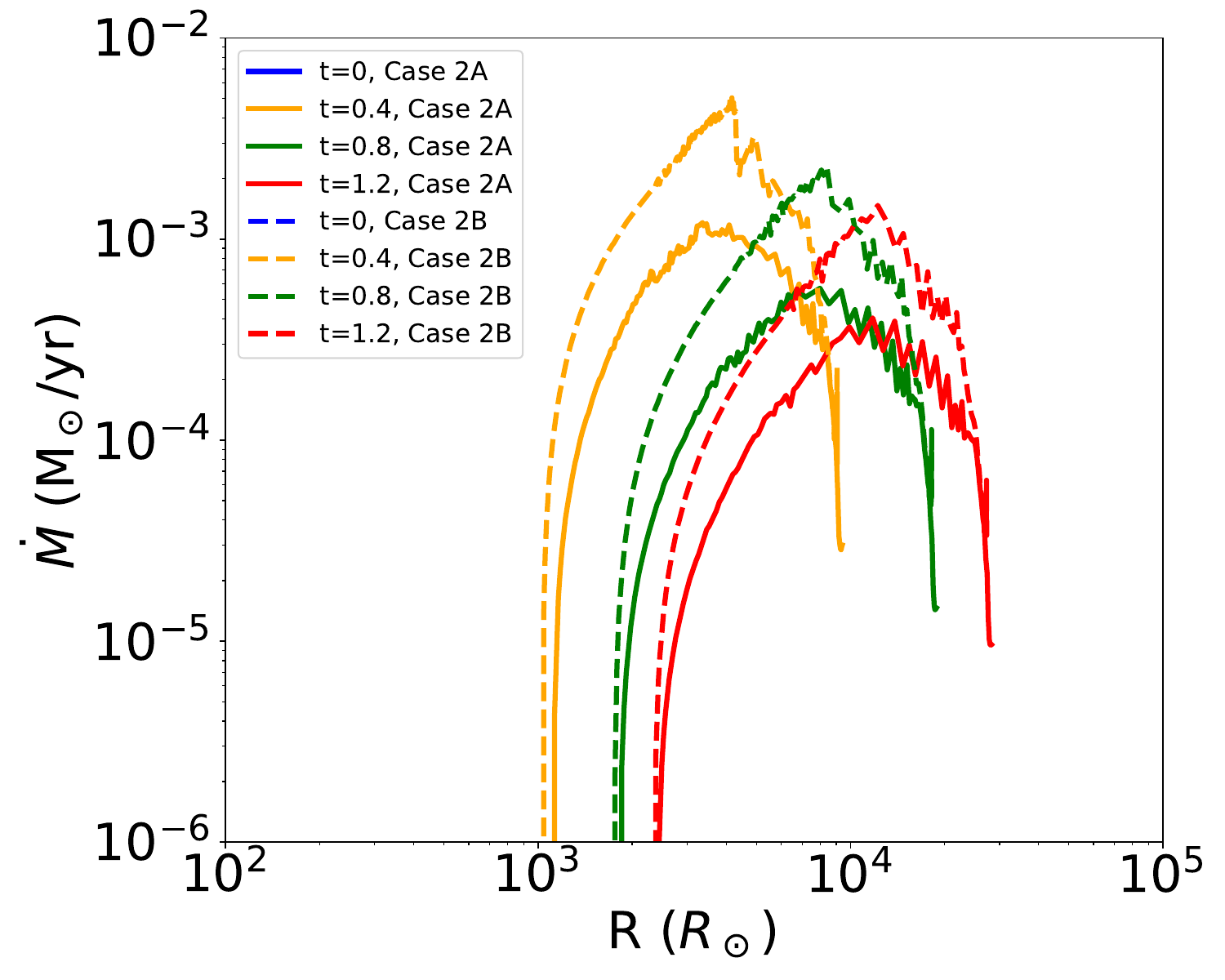}
    
    &
     \hspace{-0.9cm} 
    \includegraphics[trim={0.2cm 0.0cm 0.0cm 0.2cm},clip,width=0.5\textwidth]{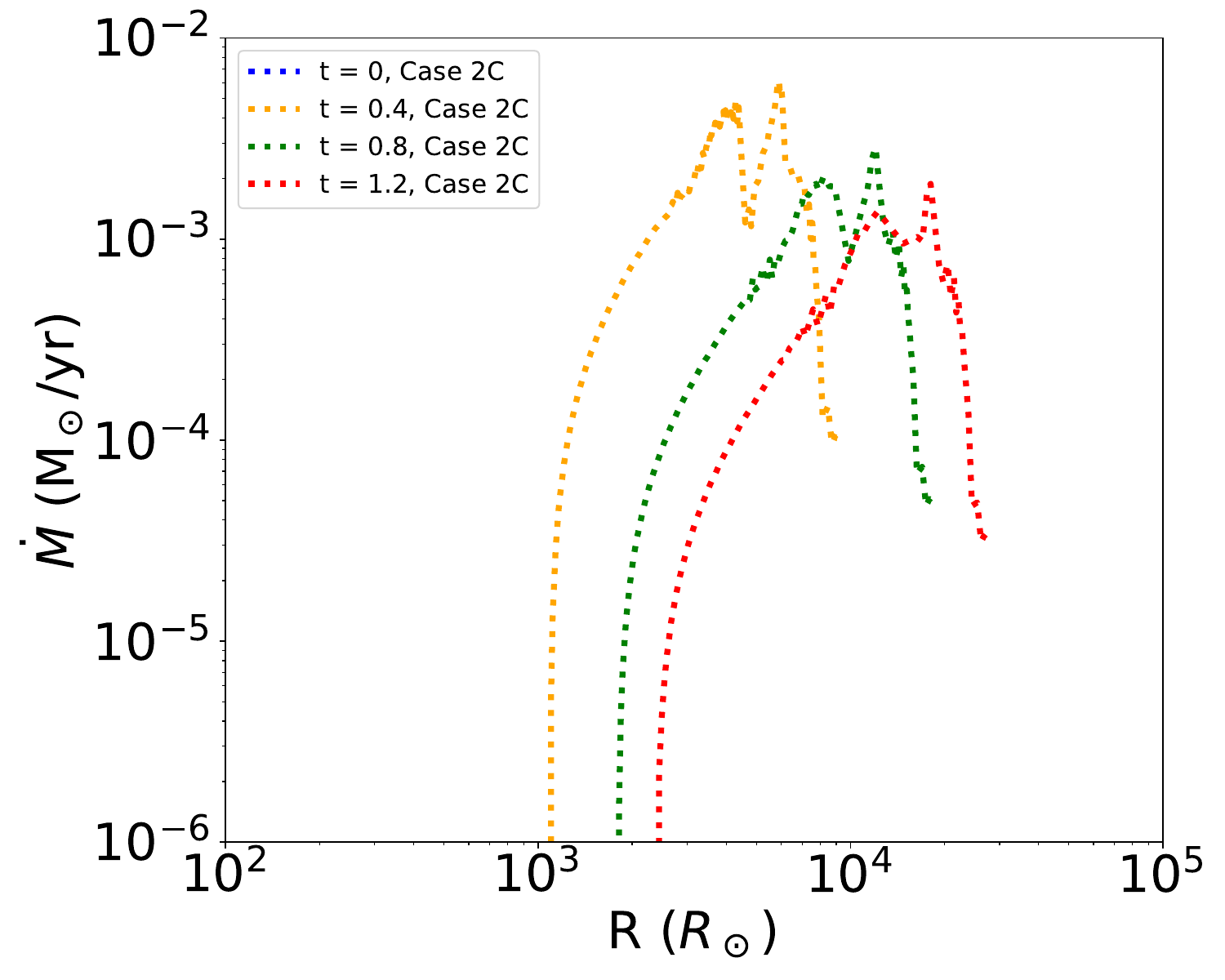}
    \\
    \small (a) & (b)
  \end{tabular}
  \caption{Panel (a): Mass-loss rate profiles, calculated using $\dot{M}(r) = 4\pi r^2 \rho (r) v (r)$, at $t = 0$, 0.4, 0.8, and 1.2 years for Case 2A, and Case 2B. Here the solid lines correspond to Case 2A (section \ref{3.2}), for the width of the heating region of 0.5 $\rm R_{\odot}$, while the dashed region corresponds to Case 2B, where the width of the heating 0.8 $\rm R_{\odot}$ (section \ref{3.3}). Panel (b): Mass-loss rate profiles, calculated using $\dot{M}(r) = 4\pi r^2 \rho (r) v (r)$, at $t = 0$, 0.4, 0.8, and 1.2 years for Case 2C for heating region 0.5 $\rm R_{\odot}$ (section \ref{3.4}). }
  \label{fig4}
  \end{figure*}

In this case, the heating source is present within the region between 23.0 to 23.5 $ \rm R_{\odot}$, with an energy deposition value mentioned in Table~\ref{T2} for all three time periods $t = 0.4, 0.8, 1.2$ years. As a result of the energy deposition, the stellar envelope begins to expand outward. The stellar radius initially, $t = 0 $ years, is 90.36 $\rm R_{\odot}$, and after 1.2 years, it expands to $R_{\rm ext}$ = $\approx 10^{4}\, \rm R_{\odot}$ (stellar radius + outflow) i.e., by the factor of 312. The increased values of the extended radius itself suggest the presence of the outflow. As shown in Figure \ref{fig3} panel (b), the total energy in the outer layers of the star becomes positive following the energy deposition, indicating the initiation of an outflow. This suggests that the injected energy is sufficient to unbind the stellar material, leading to the onset of an outflow. Similarly, the density decreases in the outer layers and reaches up to $10^{-18}~ \rm g \,cm^{-3}$, indicating the extended envelope with the outflow region as shown in panel (b). In both panels, near \( R_{\rm out} \simeq 58.1\,R_{\odot} \), we notice this transition: the total energy becomes positive, indicating that the outer layers are no longer gravitationally bound, while the density profile extends to larger radii. We therefore identify 58.1 $\rm R_{\odot}$ as the onset radius of the outflow. Therefore, during the hydrodynamic calculations, the deposited energy results in the unbinding of material in the outer layers, leading to an outflow. By using the outflow velocity and density, we later compute the mass and the outflow rate of the material that leaves the star. First, we calculate the mass of this outflow
\begin{equation}
\label{continuity}
 \\    M_{\rm out} = \int_{R_{\rm out}}^{\infty}4 \pi r^{2}\, \rho (r)\, dr  ,
 \label{eq5}
\end{equation}
where $R_{\rm out} \approx 58.1 \, \rm R_{\odot}$ represents the radius at which the material begins to unbind, and $\rho$ shows the density of the unbound material. Thus, the unbind mass ($M_{\rm out}$) at 
$t = 0.4, 0.8, 1.2$ years, are 0.0006, 0.00054, 0.00051 $\rm M_{\odot}$ respectively. Thus, the average outflow rate is
\begin{equation}
    \dot{M} = \frac{M_{\rm out}(t_{f}) - M_{\rm out}(t_{i})}{t_{f} - t_{i}}.
    \label{eq6}
\end{equation}

As shown in Figure \ref{fig4} panel (a), during Case 1, the absence of the outflow due to hydrostatic equilibrium results in no outflow (i.e. there is no outflow rate profile in the Figure for Case 1). However, in Case 2A, there is the presence of an outflow, and we observe the profiles of outflow for Case 2A, which are shown by solid lines in Figure~\ref{fig4} panel (a).
Using the above values of unbind mass, the analytical values of the outflow rate, at $t=0.4, 0.8, 1.2$ years, is $ 1.5, 0.7, 0.45 \times 10^{-3}\, \rm M_{\odot}\, yr^{-1}$ respectively. As seen in the Figure \ref{fig4} panel (a), the outflow at $t = 0.4$ years shows a higher peak compared to later times. Because the energy is deposited in the form of a thermal bomb, which triggers an immediate and strong outflow before the star has time to expand significantly. As the star expands over time, the outflow weakens, resulting in a lower outflow rate at later stages. It is also important to note that the outflow profiles are instantaneous results of the energy deposition, showing sharp peaks with values exceeding \(\rm 10^{-3}\,M_{\odot}\,yr^{-1}\). In contrast, the outflow rates reported above are analytically estimated by considering the total unbound material and the corresponding time interval. These analytical values, therefore, represent an average outflow rate over the entire ouflow episode. As a result, the instantaneous outflow profiles exhibit higher peak values than the analytically derived averages, as shown in Figure~\ref{fig4} panel (a).

Since the artificial energy deposition leads to the unbinding of stellar material, it results in the formation of an outflow. The onset of the outflow is identified by examining the total energy profiles, specifically, the outflow radial point at which the total energy becomes positive (i.e., $R_{\rm out} = 58.1\, \rm R_{\odot}$). As shown in Figure~\ref{fig3}, panel (b), the region where energy is deposited corresponds to the location where the total energy begins to rise above zero. In this model, energy was injected between 23 and 23.5~$\rm R_\odot$, and the outflow radius is at approximately 58.1~$\rm R_\odot$, where the stellar material first becomes unbound. This marks the radial point beyond which material is no longer gravitationally bound and can escape the star as an outflow. To quantify the energetic properties of the outflow, we calculate the $E_{\rm k}$ carried by the unbound material
\begin{equation}
E_{\text{k, shell}} = \frac{1}{2} \, m_{\text{shell}} \, v_{\text{shell}}^2,
\label{eq7}
\end{equation}
where, $m_{\rm shell}$ is the mass of the shell, and $v_{\rm shell} $ is its velocity. By summing the $E_{\rm k}$ of all unbound shells from the stellar radius to the outer boundary of the model, we obtain the total $E_{\rm k}$ carried by the outflow. At $t = 0.4,\, 0.8,\, \text{and}\, 1.2~\text{yr}$, the $E_{\rm k}$ released by the outflow is approximately $1.2,\, 1.16,\, \text{and}\, 1.08 \times 10^{47}\,\mathrm{erg}$, respectively. The corresponding outflow velocities at these times are $4.5,\, 4.6,\, \text{and}\, 4.7 \times 10^{3}\,\mathrm{km\,s^{-1}}$, respectively. By comparing these values with the deposited energy, we estimate the fraction of the deposited energy that is carried away by the ejected material. At $t = 0.4~\text{yr}$, approximately $24\%$ of the deposited energy is ejected, while at $t = 0.8~\text{yr}$ this fraction decreases to about $12\%$. At $t = 1.2~\text{yr}$, the fraction becomes negligible, around $0.07\%$. All other corresponding stellar parameters and their fraction are given in Table~\ref{T3}.  It is important to note that the deposited energy, \(E_{\mathrm{dep}}\), is lower than the energy at which the binding and gravitational energies intersect (see Figure~\ref{fig_1}). Consequently, depositing the energy at \(R_{\rm crit} \simeq 23\,\rm R_{\odot}\) does not unbind material, and we do not observe outflow from this radius. Instead, outflow appears near \(R_{\rm out} \simeq 58.1\,\rm R_{\odot}\), where \(E_{\mathrm{dep}}\) exceeds the local binding energy. The chosen values are guided by the \textsc{mesa} calculations: higher \(E_{\mathrm{dep}}\) values tend to cause convergence issues, which we deliberately avoid to ensure numerical stability.

\begin{table*}
\caption{Summary of the outflow properties derived for different cases at three characteristic times ($t = 0.4$, $0.8$, and $1.2~\rm yr$). 
The table lists the total ejected mass ($M_{\rm out}$), $E_{\rm k}$ of the outflow ($E_{\rm k, out}$), outflow rate ($\dot{M}$), outflow velocity ($V_{\rm out}$), and the fractional energy ejected relative to the deposition energy. 
The variations in luminosity ($\Delta L$) and its factor, effective temperature ($\Delta T_{\rm eff}$) and its factor. Here $\Delta R$ denotes the difference between the initial stellar radius (90.36 $\rm R_{\odot}$) before energy deposition, and the final stellar radius ($R_{\rm ext}$)
after the star expands in response to the deposited energy.
and its expansion factor corresponds to the response to energy deposition. Case~1 represents the hydrostatic reference model without outflow. A plus sign denotes an increase; a minus sign denotes a decrease.}
\centering
\begin{tabular}{lcccccccccccc}
\hline\hline
Case & $M_{\rm out}$ & $E_{\rm k, out}$ & $\dot M$ ($10^{-3}\,$ & $V_{\rm out}$ & Fraction & $\Delta L$ & $f_L$ & $\Delta T_{\rm eff}$ & $f_{T_{\rm eff}}$ & $\Delta R$ & $f_R$ \\
 & ($M_\odot$) & ($10^{47}\,\rm{erg}$) & $\rm M_\odot\,\rm{yr}^{-1}$) & (km\,s$^{-1}$) & (\%) & ($10^{5}\,\rm L_\odot$) &  & ($10^{3}\,\rm{K}$) &  & ($\rm R_\odot$) &  \\
\hline\hline
\multicolumn{12}{l}{\textbf{Case 1}}\\
$t = 1.2$ & 0 & 0 & 0 & 0 & -- & $-3.43$ & $-0.71$ & $-2.5$ & $-0.78$ & $+8.22$ & $+1.09$ \\
\hline
\multicolumn{12}{l}{\textbf{Case 2A}}\\
$t = 0.4$ & 0.0006 & 1.2  & 1.5  & 4.5 & 24   & $+3.10$ & $+1.26$ & $-17.83$ & $-0.10$ & $+9320$            & $+109$ \\
$t = 0.8$ & 0.0005 & 1.16 & 0.7  & 4.6 & 12   & $+2.99$ & $+1.26$ & $-18.4$  & $-0.07$ & $+1.88\times10^{4}$ & $+208$ \\
$t = 1.2$ & 0.0005 & 1.08 & 0.45 & 4.7 & 0.07 & $+2.97$ & $+1.26$ & $-18.7$  & $-0.06$ & $+2.82\times10^{4}$ & $+312$ \\
\hline
\multicolumn{12}{l}{\textbf{Case 2B}}\\
$t = 0.4$ & 0.0017 & 3.90 & 4.45 & 4.7 & 81   & $+3.86$ & $+1.33$ & $-17.8$  & $-0.10$ & $+8990$            & $+100$ \\
$t = 0.8$ & 0.0016 & 3.57 & 2    & 4.6 & 37   & $+3.22$ & $+1.28$ & $-18.4$  & $-0.08$ & $+1.82\times10^{4}$ & $+201$ \\
$t = 1.2$ & 0.0015 & 3.38 & 1.3  & 4.6 & 0.23 & $+3.02$ & $+1.26$ & $-18.7$  & $-0.06$ & $+2.72\times10^{4}$ & $+301$ \\
\hline
\multicolumn{12}{l}{\textbf{Case 2C}}\\
$t = 0.4$ & 0.0015 & 3.44 & 3.75 & 4.8 & 47   & $+3.02$ & $+1.26$ & $-17.7$  & $-0.11$ & $+8901$            & $+99.32$ \\
$t = 0.8$ & 0.0014 & 3.08 & 1.7  & 4.6 & 21   & $+2.85$ & $+1.24$ & $-18.4$  & $-0.08$ & $+1.79\times10^{4}$ & $+199$ \\
$t = 1.2$ & 0.0013 & 3.07 & 1.15 & 4.7 & 14   & $+2.79$ & $+1.23$ & $-18.6$  & $-0.06$ & $+2.69\times10^{4}$ & $+298$ \\
\hline
\end{tabular}
\label{T3}

\end{table*}

\subsection{Case 2B: Effect of the heating-region width on the numerical redistribution of energy during the hydrodynamical case}
\label{3.3}

To investigate how the outflow profiles depend on the width of the heating region, we now deposit the energy over a broader region in the stellar interior, i.e. $\delta R = 0.8 \, \rm R_{\odot}$ (Case 2B). While the previous Case 2A involved energy deposition between 23 and 23.5 $\rm R_{\odot}$, in this setup the deposition region is extended from 23 to 23.8 $\rm R_{\odot}$. The values of energy deposition, and width of the heating region are also mentioned in Table~\ref{T2}.  Thus, this section~\ref{3.3} is the extended version of Case 2A, by varying the width of the heating region.
This modification allows us to understand how increasing the width of the heating region influences the resulting outflow and the outflow-driving efficiency. At $t=0$ years, the stellar radius is approximately 90.36 $\rm R_{\odot}$. Following the energy deposition, the stellar radius expands, reaching about $R_{\rm exten}$ = $90.54 \, \rm R_{\odot}$. By analyzing the total energy profiles, we find that increasing the width of the energy deposition region systematically shifts the onset of the outflow, at \( R_{\rm out} = 60.3 \, \rm R_{\odot} \), in this case. For broader heating regions, the onset of positive total energy, and hence the initiation of the outflow, is pushed outward. This behavior is expected: as the heating region expands, more outer layers receive energy input, resulting in a larger radial zone where the total energy becomes positive. Consequently, the outflow is launched from progressively larger radii, marking the point at which the stellar material transitions from a bound to an unbound state. During this expansion phase, the luminosity increases, while the effective temperature decreases. The variations of all stellar parameters, along with their respective factors, are presented in Table~\ref{T3}. In comparison, where the energy is deposited in a narrower region Case 2A (see section \ref{3.2}), the star expands slightly more. These results suggest that a narrower (more localised) heating region is more efficient at driving expansion to larger radii and achieving slightly higher luminosities, while broader heating regions result in slightly less expansion. This reflects how the spatial distribution of deposited energy influences the star’s structural and thermal responses.

Furthermore, using Equations~\ref{eq5} and \ref{eq6}, we also calculated the unbound mass of the outflow and the corresponding outflow rate profile in this broader heating region. In this scenario, at $t = 0.4$, $0.8$, and $1.2$ years, the unbound mass values are approximately $0.00178$, $0.00164$, and $0.00158~\rm M_{\odot}$, respectively.  At $t = 0.4~\text{yr}$, approximately $81\%$ of the deposited energy is ejected, while at $t = 0.8~\text{yr}$ this fraction decreases to about $37\%$. At $t = 1.2~\text{yr}$, the fraction becomes negligible, around $0.23\%$. All corresponding values of $E_{\rm k}$ and outflow velocities are summarised in Table~\ref{T3}.
 The values of the outflow rate in this case indicate that both the total unbound mass and the outflow rate are higher when the heating region is broader, compared to a narrower region. This trend is also evident in the outflow rate profile shown in Figure~\ref{fig4} panel (a), where the dashed lines illustrate that increasing the width of the heating region leads to a significant enhancement in outflow. This behavior arises because broader heating deposits energy across a larger mass range within the stellar envelope, energizing more of the outer layers and enabling a greater fraction of mass to participate in the outflow. The distributed energy input also results in a more gradual pressure gradient, which, while slightly less effective at accelerating material to extremely large radii, drives a more sustained and mass-loaded outflow over time. 
In contrast, when the heating region is narrower (as in the localized energy deposition case, section~\ref{3.2}), the same amount of energy is deposited into a smaller mass. This creates a steeper pressure gradient, resulting in a more impulsive acceleration of material. While this produces a rapid and extended initial expansion, the outflow carries away less total mass, and the outflow rate declines more quickly as the localized energy is exhausted and the star expands.
Thus, in the broader heating case, although the star expands to a slightly lower radius, the energy deposition is more evenly distributed, allowing a larger portion of the envelope to contribute to the outflow. This results in a higher total unbound mass and a more sustained outflow rate over a longer period. In essence, broader heating enhances the efficiency of the outflow, while narrower heating enhances the efficiency of expansion.
The $E_{\rm k}$ values are notably higher, as shown in Table~\ref{T3}, than those obtained for the narrower heating region. This is expected, as the broader energy deposition not only energizes a larger portion of the envelope but also results in a more massive and sustained outflow. Consequently, more material is accelerated to significant velocities, increasing the overall $E_{\rm k}$ carried by the outflow. This further supports the conclusion that broader heating regions are more effective at driving strong, massive outflow, even if the stellar expansion is somewhat less extreme than in the narrow deposition case.

\subsection{Case 2C: Hydrodynamic response for the higher $E_{\rm dep}$}
\label{3.4}

To investigate how the outflow profile depends on the deposited energy ($E_{\rm dep}$), we vary the value of $E_{\rm dep}$ into the stellar interior by a factor of 1.5 i.e., Case 2C as compare to Case 2A, while keeping the injection region fixed between 23 and 23.5 $\rm R_{\odot}$. The value of energy deposition for each time period is mentioned in Table~\ref{T2}.

Thus, this section~\ref{3.4}, is the extended version of Case 2A, by varying the $E_{\rm dep}$. This setup allows us to examine how increasing $E_{\rm dep}$ influences both the outflow and its driving efficiency. From the total energy profiles, we observe that as $E_{\rm dep}$ increases, the outflow radius at which the material becomes unbound remains nearly constant at around 58.1 $\rm R_{\odot}$, which is consistent with the result found in section \ref{3.2}, i.e., Case 2A. At $t=0$ years, the stellar radius is approximately 90.36 $\rm R_{\odot}$. Following the energy deposition, the star undergoes an expansion, with its radius increasing to about $ R_{\rm ext} \approx 10^{4} \, \rm R_{\odot} \rm (stellar\, radius + outflow$). 
Interestingly, when a higher amount of energy is deposited in the stellar envelope, the resulting stellar configuration becomes somewhat hotter, less luminous, and less expanded compared to the Case 2A with a lower deposited energy, as discussed in section~\ref{3.2}. This counterintuitive outcome arises due to the more violent and impulsive response of the envelope to the stronger energy input. In such cases, a significant portion of the deposited energy is rapidly converted into $E_{\rm k}$ of the ejecta, and much of it is carried away by the escaping material.

Consequently, less energy remains available to drive a prolonged expansion of the envelope, leading to a more compact, hotter, and less luminous post-outburst configuration. In our models, the rapid outflow of material efficiently removes energy from the system, limiting further inflation of the envelope and reducing the emergent luminosity despite the high initial energy input. This behavior highlights the complex and non-linear hydrodynamic response of the stellar envelope to varying levels of deposited energy. In this case, the higher energy leads to a more impulsive and rapid outflow. As a result, the amount of unbound mass within the outflow in this case is 0.0015, 0.0014, and 0.0013 $\rm M_{\odot}$ at $ t = 0.4, 0.8, 1.2$ respectively. The outflow rate values by using Equations \ref{eq5} and \ref{eq6} are given in Table~\ref{T3}, along with the $E_{\rm k}$ and velocity values carried out by the outflow. At $t = 0.4~\text{yr}$, approximately $47\%$ of the deposited energy is ejected, while at $t = 0.8~\text{yr}$ this fraction decreases to about $21\%$. At $t = 1.2~\text{yr}$, the fraction is around $14\%$. The variation in the stellar parameters and their factor are given in Table~\ref{T3} for this case as well. Figure~\ref{fig4} panel (b) presents the outflow rate profiles for this scenario. The displayed profiles represent the integrated values, calculated from the stellar radius outward to infinity. While the outflow rate values given in Table~\ref{T3} are derived analytically using Equation~\ref{eq6} time interval is 0.4 years. 

\section{Discusion}
\label{4}

Our numerical experiments show that energy deposition within the envelope of a massive star produces a strongly non-linear hydrodynamic response, and as a result, it initiates the outflow. Under the certain condition of the electron scattering, after the deposition of the energy, the envelope undergoes significant inflation and transitions from hydrostatic expansion to an outflow. The onset, strength, and temporal structure of this outflow depend on both the magnitude of the deposited energy and the width of the heating region. In particular, we find threshold behavior in the emergence of outflow and systematic trends whereby higher deposited energy leads to more rapid energy removal through expelled material, limiting prolonged envelope expansion. These outflows are of particular interest in evolutionary phases where intense energy is deposited into the stellar interior, potentially initiating strong outflows.

The connection to these observed eruptive phenomena in massive stars must, however, be treated with caution. Although LBV and pre-supernova outbursts provide a natural point of comparison, the present simulations differ from these events in several key respects. The outflow phase in our models are short-lived, the total ejected masses are modest compared to those for LBV outflows, and the characteristic outflow velocities are much lower than observed in LBVs. These differences strongly limit a direct, quantitative application of our results to LBV outbursts. Our models do not attempt to reproduce the physical trigger, duration, or energetics of such events.

In Case 1, after depositing the energy, the star’s envelope does not immediately respond with the outflow. Instead, it undergoes quasi-hydrostatic expansion to larger radii, as illustrated in Figure \ref{fig2} panel (b). In contrast, in Case 2A, when conditions become favorable (see section~\ref{3.2} for details), there is the emergence of an outflow. As the deposited energy is attributed to unbinding the material in the outer layers. During this Case 2A, there is an outflow of the material. These results indicate that radiation diffusion plays a minimal role in regulating the outflow’s behavior when the $E_{\rm k}$ being carried exceeds or matches the Eddington luminosity. Although the physical setup and methodology apply to a broad class of astrophysical scenarios, we focus here on massive star outflow by depositing the energy, instead of comparing it to any other astrophysical phenomena. The duration of these outflows, far exceeding dynamical timescales, suggests they are not driven by explosive shocks, but rather by a sustained energy source capable of powering a continuous outflow. Although the physical mechanism responsible for such energy release remains uncertain, our simulations demonstrate that localized energy deposition can generate strong outflows with mass-loss rates and velocities of several hundred $\rm km~s^{-1}$, values that lie in the range encountered in a variety of energetic astrophysical environments. We emphasize that our models are not intended to reproduce any particular class of transients. Rather, they are designed to isolate the envelope’s hydrodynamic response to imposed energy input and to characterize the resulting outflow properties in this, idealized framework. We also explore the double-peaked outflow behavior of the star during the higher deposition of energy in Case 2C.  As shown in Figure~\ref{fig4} panel (b), during this time,  with higher energy deposition (section \ref{3.3}), the outflow profile exhibits two distinct peaks in the outflow. In contrast, the profiles presented in Figure~\ref{fig4}, corresponding to the lower energy deposition and broader width cases (i.e., Case 2A and Case 2B, respectively) discussed in sections \ref{3.2} and \ref{3.3}, do not display such double-peaked behavior. This suggests that the appearance of a double-peak structure in the outflow is closely associated with higher energy input. The occurrence of two outflow peaks implies that the envelope may undergo multiple ejection episodes during the outburst event. Therefore, such behavior could serve as a potential signature in observations, indicating repeated outbursts or multi-peaked light curves in massive stars.

\section{Summary}
\label{5}
We examine the stellar response to the deposition of the energy originated from a binary companion. We compare the deposited energy to the orbital gravitational energy that would be released if a companion were to spiral into a separation where this energy becomes comparable to the envelope binding energy. The simulations themselves, however, follow the evolution of a single star, and no binary dynamics are included. The imposed energy deposition should therefore be interpreted as a parameterized representation of processes such as tidal heating, rather than as a self-consistent model of binary evolution. The location of energy deposition is determined based on calculations of gravitational and binding energies. The energy is deposited at the location of the star where these two energies become equal (Figure \ref{fig_1}).
Case 1, representing the hydrostatic scenario, we include the inversions in gas pressure and density within the stellar envelope. These structural instabilities result in the formation of an extended envelope. As a consequence, rather than driving an outflow, the deposited energy goes primarily to \textit{inflate the envelope to much larger radii}. This expansion leads to a decrease in luminosity and effective temperature, as discussed in section \ref{3.1}. 

In Case 2A, which represents the hydrodynamic scenario, the star is allowed to respond dynamically to the deposition of the energy. Instead of developing pressure and density inversions as in the hydrostatic case, \textit{the stellar envelope becomes unstable and initiates an outflow}, which transports both energy and mass away from the star as shown in Section \ref{3.2}. The onset of this outflow marks a shift from a quasi-stable envelope structure to one dominated by outflow and dynamical evolution, during which the star’s luminosity increases while its effective temperature decreases. These changes are significantly more pronounced compared to the hydrostatic case. In the hydrostatic simulation (Case 1), no outflow develops, and despite the formation of an extended envelope, the star remains relatively compact. In contrast, the hydrodynamic simulation (Case 2A) results in the launch of a outflow that continues to propagate outward. In our simulations, the \textsc{mesa} profiles do not show any explicit outflow, as the stellar mass remains conserved throughout the simulation period. Instead, we estimate the outflow rate based on the amount of material that becomes unbound due to the deposited energy, as described in Section~\ref{3.2} (equations \ref{eq5} and \ref{eq6}). As illustrated in Figure \ref{fig4}, the outflow rate in the hydrodynamic case reaches up to $\rm 10^{-2}~M_{\odot}\, yr^{-1}$.

We also investigate how the outflow profile depends on the width of the heating region (Case 2B) and the magnitude of the deposited energy (Case 2C). By varying the extent of the heating layer and increasing the deposited energy, we find that a broader heating region leads to a higher mass-loss rate and shifts the unbinding radius outward. Similarly, the variation in luminosity is higher in Case 2B, compared to Case 2A. While higher deposited energy triggers higher outflow rates, it results in a somewhat hotter and less luminous star due to the rapid loss of energy through expelled material.

Once a star experiences such an outflow event, the remaining envelope expands, and its density profile is significantly altered. As a result, if another abrupt energy injection occurs, the properties of the eruption will not resemble the previous one. This is consistent with the findings of our study \citep{MukhijaKashi2025}, in which we simulate three successive giant eruptions in a 100 M$_\odot$ star. In that work, we show that multiple giant eruptions (MGEs) have a cumulative effect: each eruption leaves the star in a different thermal and structural state, altering its luminosity, temperature, and recovery timescale. After each eruption, the luminosity initially decreases and then rises again as the star approaches equilibrium, with recovery occurring more rapidly at lower metallicity. At higher mass-loss rates, the outer layers exhibit oscillations and compression, signatures of thermal imbalance, that may create conditions favorable for a subsequent eruption. These results demonstrate that repeated outflow phase fundamentally change the envelope structure, ensuring that each outflow differs from the last.

Our simulations are not designed to reproduce any specific transient or evolutionary phase of real stars.
Consequently, we do not attempt to match our predicted luminosity increases, ejected masses, or outflow durations to those of LBV eruptions, pre-supernova outbursts, or other transients. Our goal is instead to characterize limiting envelope responses and outflow behaviors in a controlled framework. In s future work, we plan to extend this study to a systematic grid of stellar models, spanning a broader range of initial masses, evolutionary stages, and deposition parameters. Such an approach will enable us to determine how stellar mass and structure influence the threshold energy required to trigger different classes of outflow and to assess how these theoretical trends may qualitatively relate to the diversity of variability and outflow behavior observed in massive stars in real.

\vspace{0.5cm}

We acknowledge the Ariel HPC Center at Ariel University for providing computing resources that have contributed to the research results reported in this paper. BM acknowledges support form the AGASS center at Ariel University. The \textsc{mesa} inlists and input files to reproduce our simulations and associated data products are available on Zenodo (10.5281/zenodo.15577892).
\newpage

\bibliography{Ref}

@article{MukhijaKashi2025,
  author       = {Mukhija, B. and Kashi, A.},
  title        = {Pre-Supernova Multiple Giant Eruptions in Massive Stars},
  journal      = {ApJ},
  year         = {2025},
  note         = {Accepted},
}

@ARTICLE{mukhija2025accretionrecoverygianteruptions,
       author = {{Mukhija}, Bhawna and {Kashi}, Amit},
        title = "{Accretion and Recovery in Giant Eruptions of Massive Stars}",
      journal = {\apj},
         year = 2025,
        month = jun,
       volume = {986},
       number = {2},
          eid = {188},
        pages = {188},
          doi = {10.3847/1538-4357/add3f1},
archivePrefix = {arXiv},
       eprint = {2504.19884},
 primaryClass = {astro-ph.SR},
       adsurl = {https://ui.adsabs.harvard.edu/abs/2025ApJ...986..188M}
}

@ARTICLE{2025PASP..137j4201M,
       author = {{Mukhija}, Bhawna and {Kashi}, Amit},
        title = "{Wind Accretion in Massive Binaries Experiencing High Mass Loss Rates. I. Dependency on Mass Ratio and Orbital Period}",
      journal = {\pasp},
         year = 2025,
        month = oct,
       volume = {137},
       number = {10},
          eid = {104201},
        pages = {104201},
          doi = {10.1088/1538-3873/ae0789},
archivePrefix = {arXiv},
       eprint = {2509.12725},
 primaryClass = {astro-ph.SR},
       adsurl = {https://ui.adsabs.harvard.edu/abs/2025PASP..137j4201M}
}

@ARTICLE{2026NewA..12202475M,
       author = {{Mukhija}, Bhawna and {Kashi}, Amit},
        title = "{High power accretion in massive binary systems and the impact of metallicity}",
      journal = {\na},
         year = 2026,
        month = jan,
       volume = {122},
          eid = {102475},
        pages = {102475},
          doi = {10.1016/j.newast.2025.102475},
archivePrefix = {arXiv},
       eprint = {2509.10002},
 primaryClass = {astro-ph.SR},
       adsurl = {https://ui.adsabs.harvard.edu/abs/2026NewA..12202475M}
}

@ARTICLE{2022ApJ...924...15J,
       author = {{Jacobson-Gal{\'a}n}, W.~V. and {Dessart}, L. and {Jones}, D.~O. and {Margutti}, R. and {Coppejans}, D.~L. and {Dimitriadis}, G. and {Foley}, R.~J. and {Kilpatrick}, C.~D. and {Matthews}, D.~J. and {Rest}, S. and {Terreran}, G. and {Aleo}, P.~D. and {Auchettl}, K. and {Blanchard}, P.~K. and {Coulter}, D.~A. and {Davis}, K.~W. and {de Boer}, T.~J.~L. and {DeMarchi}, L. and {Drout}, M.~R. and {Earl}, N. and {Gagliano}, A. and {Gall}, C. and {Hjorth}, J. and {Huber}, M.~E. and {Ibik}, A.~L. and {Milisavljevic}, D. and {Pan}, Y. -C. and {Rest}, A. and {Ridden-Harper}, R. and {Rojas-Bravo}, C. and {Siebert}, M.~R. and {Smith}, K.~W. and {Taggart}, K. and {Tinyanont}, S. and {Wang}, Q. and {Zenati}, Y.},
        title = "{Final Moments. I. Precursor Emission, Envelope Inflation, and Enhanced Mass Loss Preceding the Luminous Type II Supernova 2020tlf}",
      journal = {\apj},
         year = 2022,
        month = jan,
       volume = {924},
       number = {1},
          eid = {15},
        pages = {15},
          doi = {10.3847/1538-4357/ac3f3a},
archivePrefix = {arXiv},
       eprint = {2109.12136},
 primaryClass = {astro-ph.HE},
       adsurl = {https://ui.adsabs.harvard.edu/abs/2022ApJ...924...15J}
}

@ARTICLE{2024A&A...686A.231R,
       author = {{Reguitti}, A. and {Pignata}, G. and {Pastorello}, A. and {Dastidar}, R. and {Reichart}, D.~E. and {Haislip}, J.~B. and {Kouprianov}, V.~V.},
        title = "{Searching for precursor activity of Type IIn supernovae}",
      journal = {\aap},
         year = 2024,
        month = jun,
       volume = {686},
          eid = {A231},
        pages = {A231},
          doi = {10.1051/0004-6361/202348679},
archivePrefix = {arXiv},
       eprint = {2403.10398},
 primaryClass = {astro-ph.HE},
       adsurl = {https://ui.adsabs.harvard.edu/abs/2024A&A...686A.231R}
}

@ARTICLE{2024ApJ...976..125D,
       author = {{Deman}, Julian A. and {Oey}, M.~S.},
        title = "{Kinematic Insights into Luminous Blue Variables and B[e] Supergiants}",
      journal = {\apj},
         year = 2024,
        month = nov,
       volume = {976},
       number = {1},
          eid = {125},
        pages = {125},
          doi = {10.3847/1538-4357/ad8134},
archivePrefix = {arXiv},
       eprint = {2410.06448},
 primaryClass = {astro-ph.SR},
       adsurl = {https://ui.adsabs.harvard.edu/abs/2024ApJ...976..125D}
}

@ARTICLE{2025AJ....169..128S,
       author = {{Spejcher}, Becca and {Richardson}, Noel D. and {Pablo}, Herbert and {Beltran}, Marina and {Butler}, Payton and {Avila}, Eddie},
        title = "{An Investigation into the Variability of Luminous Blue Variable Stars with TESS}",
      journal = {\aj},
         year = 2025,
        month = mar,
       volume = {169},
       number = {3},
          eid = {128},
        pages = {128},
          doi = {10.3847/1538-3881/ada561},
archivePrefix = {arXiv},
       eprint = {2501.00240},
 primaryClass = {astro-ph.SR},
       adsurl = {https://ui.adsabs.harvard.edu/abs/2025AJ....169..128S}
}

@ARTICLE{2024ApJ...974..270C,
       author = {{Cheng}, Shelley J. and {Goldberg}, Jared A. and {Cantiello}, Matteo and {Bauer}, Evan B. and {Renzo}, Mathieu and {Conroy}, Charlie},
        title = "{A Model for Eruptive Mass Loss in Massive Stars}",
      journal = {\apj},
         year = 2024,
        month = oct,
       volume = {974},
       number = {2},
          eid = {270},
        pages = {270},
          doi = {10.3847/1538-4357/ad701e},
archivePrefix = {arXiv},
       eprint = {2405.12274},
 primaryClass = {astro-ph.SR},
       adsurl = {https://ui.adsabs.harvard.edu/abs/2024ApJ...974..270C}
}

@ARTICLE{2018Natur.561..498J,
       author = {{Jiang}, Yan-Fei and {Cantiello}, Matteo and {Bildsten}, Lars and {Quataert}, Eliot and {Blaes}, Omer and {Stone}, James},
        title = "{Outbursts of luminous blue variable stars from variations in the helium opacity}",
      journal = {\nat},
         year = 2018,
        month = sep,
       volume = {561},
       number = {7724},
        pages = {498-501},
          doi = {10.1038/s41586-018-0525-0},
archivePrefix = {arXiv},
       eprint = {1809.10187},
 primaryClass = {astro-ph.SR},
       adsurl = {https://ui.adsabs.harvard.edu/abs/2018Natur.561..498J}
}

@ARTICLE{2023A&A...678A..55R,
       author = {{Rizzo}, J.~R. and {Bordiu}, C. and {Buemi}, C. and {Leto}, P. and {Ingallinera}, A. and {Bufano}, F. and {Umana}, G. and {Cerrigone}, L. and {Trigilio}, C.},
        title = "{The rich molecular environment of the luminous blue variable star AFGL 2298}",
      journal = {\aap},
         year = 2023,
        month = oct,
       volume = {678},
          eid = {A55},
        pages = {A55},
          doi = {10.1051/0004-6361/202346980},
archivePrefix = {arXiv},
       eprint = {2307.11851},
 primaryClass = {astro-ph.SR},
       adsurl = {https://ui.adsabs.harvard.edu/abs/2023A&A...678A..55R}
}

@ARTICLE{1,
       author = {{Hubble}, Edwin and {Sandage}, Allan},
        title = "{The Brightest Variable Stars in Extragalactic Nebulae. I. M31 and M33.}",
      journal = {\apj},
         year = 1953,
        month = nov,
       volume = {118},
        pages = {353},
          doi = {10.1086/145764},
       adsurl = {https://ui.adsabs.harvard.edu/abs/1953ApJ...118..353H}
}

@ARTICLE{1994PASP..106.1025H,
       author = {{Humphreys}, Roberta M. and {Davidson}, Kris},
        title = "{The Luminous Blue Variables: Astrophysical Geysers}",
      journal = {\pasp},
         year = 1994,
        month = oct,
       volume = {106},
        pages = {1025},
          doi = {10.1086/133478},
       adsurl = {https://ui.adsabs.harvard.edu/abs/1994PASP..106.1025H}
}

@INPROCEEDINGS{1999ASPC..192...32D,
       author = {{de Koter}, A. and {Vink}, J.~S. and {Lamers}, H.~J.~G.~L.~M.},
        title = "{Spectroscopic Dating of Very Massive Stars}",
    booktitle = {Spectrophotometric Dating of Stars and Galaxies},
         year = 1999,
       editor = {{Hubeny}, Ivan and {Heap}, Sally and {Cornett}, Robert},
       series = {Astronomical Society of the Pacific Conference Series},
       volume = {192},
        month = jan,
        pages = {32},
       adsurl = {https://ui.adsabs.harvard.edu/abs/1999ASPC..192...32D}
}

@ARTICLE{1999PASP..111.1124H,
       author = {{Humphreys}, Roberta M. and {Davidson}, Kris and {Smith}, Nathan},
        title = "{{\ensuremath{\eta}} Carinae's Second Eruption and the Light Curves of the {\ensuremath{\eta}} Carinae Variables}",
      journal = {\pasp},
         year = 1999,
        month = sep,
       volume = {111},
       number = {763},
        pages = {1124-1131},
          doi = {10.1086/316420},
       adsurl = {https://ui.adsabs.harvard.edu/abs/1999PASP..111.1124H}
}

@ARTICLE{2014ApJ...785...82S,
       author = {{Smith}, Nathan and {Arnett}, W. David},
        title = "{Preparing for an Explosion: Hydrodynamic Instabilities and Turbulence in Presupernovae}",
      journal = {\apj},
         year = 2014,
        month = apr,
       volume = {785},
       number = {2},
          eid = {82},
        pages = {82},
          doi = {10.1088/0004-637X/785/2/82},
archivePrefix = {arXiv},
       eprint = {1307.5035},
 primaryClass = {astro-ph.SR},
       adsurl = {https://ui.adsabs.harvard.edu/abs/2014ApJ...785...82S}
}

@ARTICLE{2007Natur.450..390W,
       author = {{Woosley}, S.~E. and {Blinnikov}, S. and {Heger}, Alexander},
        title = "{Pulsational pair instability as an explanation for the most luminous supernovae}",
      journal = {\nat},
         year = 2007,
        month = nov,
       volume = {450},
       number = {7168},
        pages = {390-392},
          doi = {10.1038/nature06333},
archivePrefix = {arXiv},
       eprint = {0710.3314},
 primaryClass = {astro-ph},
       adsurl = {https://ui.adsabs.harvard.edu/abs/2007Natur.450..390W}
}

@ARTICLE{2012MNRAS.423L..92Q,
       author = {{Quataert}, E. and {Shiode}, J.},
        title = "{Wave-driven mass loss in the last year of stellar evolution: setting the stage for the most luminous core-collapse supernovae}",
      journal = {\mnras},
         year = 2012,
        month = jun,
       volume = {423},
       number = {1},
        pages = {L92-L96},
          doi = {10.1111/j.1745-3933.2012.01264.x},
archivePrefix = {arXiv},
       eprint = {1202.5036},
 primaryClass = {astro-ph.SR},
       adsurl = {https://ui.adsabs.harvard.edu/abs/2012MNRAS.423L..92Q}
}

@ARTICLE{2011ApJS..192....3P,
       author = {{Paxton}, Bill and {Bildsten}, Lars and {Dotter}, Aaron and {Herwig}, Falk and {Lesaffre}, Pierre and {Timmes}, Frank},
        title = "{Modules for Experiments in Stellar Astrophysics (MESA)}",
      journal = {\apjs},
         year = 2011,
        month = jan,
       volume = {192},
       number = {1},
          eid = {3},
        pages = {3},
          doi = {10.1088/0067-0049/192/1/3},
archivePrefix = {arXiv},
       eprint = {1009.1622},
 primaryClass = {astro-ph.SR},
       adsurl = {https://ui.adsabs.harvard.edu/abs/2011ApJS..192....3P}
}

@ARTICLE{2013ApJS..208....4P,
       author = {{Paxton}, Bill and {Cantiello}, Matteo and {Arras}, Phil and {Bildsten}, Lars and {Brown}, Edward F. and {Dotter}, Aaron and {Mankovich}, Christopher and {Montgomery}, M.~H. and {Stello}, Dennis and {Timmes}, F.~X. and {Townsend}, Richard},
        title = "{Modules for Experiments in Stellar Astrophysics (MESA): Planets, Oscillations, Rotation, and Massive Stars}",
      journal = {\apjs},
         year = 2013,
        month = sep,
       volume = {208},
       number = {1},
          eid = {4},
        pages = {4},
          doi = {10.1088/0067-0049/208/1/4},
archivePrefix = {arXiv},
       eprint = {1301.0319},
 primaryClass = {astro-ph.SR},
       adsurl = {https://ui.adsabs.harvard.edu/abs/2013ApJS..208....4P}
}

@ARTICLE{2015ApJS..220...15P,
       author = {{Paxton}, Bill and {Marchant}, Pablo and {Schwab}, Josiah and {Bauer}, Evan B. and {Bildsten}, Lars and {Cantiello}, Matteo and {Dessart}, Luc and {Farmer}, R. and {Hu}, H. and {Langer}, N. and {Townsend}, R.~H.~D. and {Townsley}, Dean M. and {Timmes}, F.~X.},
        title = "{Modules for Experiments in Stellar Astrophysics (MESA): Binaries, Pulsations, and Explosions}",
      journal = {\apjs},
         year = 2015,
        month = sep,
       volume = {220},
       number = {1},
          eid = {15},
        pages = {15},
          doi = {10.1088/0067-0049/220/1/15},
archivePrefix = {arXiv},
       eprint = {1506.03146},
 primaryClass = {astro-ph.SR},
       adsurl = {https://ui.adsabs.harvard.edu/abs/2015ApJS..220...15P}
}

@ARTICLE{2018ApJS..234...34P,
       author = {{Paxton}, Bill and {Schwab}, Josiah and {Bauer}, Evan B. and {Bildsten}, Lars and {Blinnikov}, Sergei and {Duffell}, Paul and {Farmer}, R. and {Goldberg}, Jared A. and {Marchant}, Pablo and {Sorokina}, Elena and {Thoul}, Anne and {Townsend}, Richard H.~D. and {Timmes}, F.~X.},
        title = "{Modules for Experiments in Stellar Astrophysics (MESA): Convective Boundaries, Element Diffusion, and Massive Star Explosions}",
      journal = {\apjs},
         year = 2018,
        month = feb,
       volume = {234},
       number = {2},
          eid = {34},
        pages = {34},
          doi = {10.3847/1538-4365/aaa5a8},
archivePrefix = {arXiv},
       eprint = {1710.08424},
 primaryClass = {astro-ph.SR},
       adsurl = {https://ui.adsabs.harvard.edu/abs/2018ApJS..234...34P}
}

@ARTICLE{2019ApJS..243...10P,
       author = {{Paxton}, Bill and {Smolec}, R. and {Schwab}, Josiah and {Gautschy}, A. and {Bildsten}, Lars and {Cantiello}, Matteo and {Dotter}, Aaron and {Farmer}, R. and {Goldberg}, Jared A. and {Jermyn}, Adam S. and {Kanbur}, S.~M. and {Marchant}, Pablo and {Thoul}, Anne and {Townsend}, Richard H.~D. and {Wolf}, William M. and {Zhang}, Michael and {Timmes}, F.~X.},
        title = "{Modules for Experiments in Stellar Astrophysics (MESA): Pulsating Variable Stars, Rotation, Convective Boundaries, and Energy Conservation}",
      journal = {\apjs},
         year = 2019,
        month = jul,
       volume = {243},
       number = {1},
          eid = {10},
        pages = {10},
          doi = {10.3847/1538-4365/ab2241},
archivePrefix = {arXiv},
       eprint = {1903.01426},
 primaryClass = {astro-ph.SR},
       adsurl = {https://ui.adsabs.harvard.edu/abs/2019ApJS..243...10P}
}

@ARTICLE{2014ARA&A..52..487S,
       author = {{Smith}, Nathan},
        title = "{Mass Loss: Its Effect on the Evolution and Fate of High-Mass Stars}",
      journal = {\araa},
         year = 2014,
        month = aug,
       volume = {52},
        pages = {487-528},
          doi = {10.1146/annurev-astro-081913-040025},
archivePrefix = {arXiv},
       eprint = {1402.1237},
 primaryClass = {astro-ph.SR},
       adsurl = {https://ui.adsabs.harvard.edu/abs/2014ARA&A..52..487S}
}

@ARTICLE{2008MNRAS.390.1751K,
       author = {{Kashi}, Amit and {Soker}, Noam},
        title = "{The orientation of the {\ensuremath{\eta}} Carinae binary system}",
      journal = {\mnras},
         year = 2008,
        month = nov,
       volume = {390},
       number = {4},
        pages = {1751-1761},
          doi = {10.1111/j.1365-2966.2008.13883.x},
archivePrefix = {arXiv},
       eprint = {0806.4367},
 primaryClass = {astro-ph},
       adsurl = {https://ui.adsabs.harvard.edu/abs/2008MNRAS.390.1751K}
}

@ARTICLE{2016ApJ...817...66K,
       author = {{Kashi}, Amit and {Davidson}, Kris and {Humphreys}, Roberta M.},
        title = "{Recovery from Giant Eruptions in Very Massive Stars}",
      journal = {\apj},
         year = 2016,
        month = jan,
       volume = {817},
       number = {1},
          eid = {66},
        pages = {66},
          doi = {10.3847/0004-637X/817/1/66},
archivePrefix = {arXiv},
       eprint = {1510.06428},
 primaryClass = {astro-ph.SR},
       adsurl = {https://ui.adsabs.harvard.edu/abs/2016ApJ...817...66K}
}

@ARTICLE{2017RSPTA.37560268S,
       author = {{Smith}, Nathan},
        title = "{Luminous blue variables and the fates of very massive stars}",
      journal = {Philosophical Transactions of the Royal Society of London Series A},
         year = 2017,
        month = sep,
       volume = {375},
       number = {2105},
          eid = {20160268},
        pages = {20160268},
          doi = {10.1098/rsta.2016.0268},
       adsurl = {https://ui.adsabs.harvard.edu/abs/2017RSPTA.37560268S}
}

@ARTICLE{2010MNRAS.405.1924K,
       author = {{Kashi}, Amit},
        title = "{An indication for the binarity of P Cygni from its 17th century eruption}",
      journal = {\mnras},
         year = 2010,
        month = jul,
       volume = {405},
       number = {3},
        pages = {1924-1929},
          doi = {10.1111/j.1365-2966.2010.16582.x},
archivePrefix = {arXiv},
       eprint = {0912.3998},
 primaryClass = {astro-ph.SR},
       adsurl = {https://ui.adsabs.harvard.edu/abs/2010MNRAS.405.1924K}
}

@ARTICLE{2016MNRAS.458.1214Q,
       author = {{Quataert}, Eliot and {Fern{\'a}ndez}, Rodrigo and {Kasen}, Daniel and {Klion}, Hannah and {Paxton}, Bill},
        title = "{Super-Eddington stellar winds driven by near-surface energy deposition}",
      journal = {\mnras},
         year = 2016,
        month = may,
       volume = {458},
       number = {2},
        pages = {1214-1233},
          doi = {10.1093/mnras/stw365},
archivePrefix = {arXiv},
       eprint = {1509.06370},
 primaryClass = {astro-ph.SR},
       adsurl = {https://ui.adsabs.harvard.edu/abs/2016MNRAS.458.1214Q}
}

@INPROCEEDINGS{2012ASSL..384...43D,
       author = {{Davidson}, Kris},
        title = "{The Central Star: Instability and Recovery}",
    booktitle = {Eta Carinae and the Supernova Impostors},
         year = 2012,
       editor = {{Davidson}, Kris and {Humphreys}, Roberta M.},
       series = {Astrophysics and Space Science Library},
       volume = {384},
        month = jan,
        pages = {43},
          doi = {10.1007/978-1-4614-2275-4_3},
       adsurl = {https://ui.adsabs.harvard.edu/abs/2012ASSL..384...43D}
}

@ARTICLE{1997ARA&A..35....1D,
       author = {{Davidson}, Kris and {Humphreys}, Roberta M.},
        title = "{Eta Carinae and Its Environment}",
      journal = {\araa},
         year = 1997,
        month = jan,
       volume = {35},
        pages = {1-32},
          doi = {10.1146/annurev.astro.35.1.1},
       adsurl = {https://ui.adsabs.harvard.edu/abs/1997ARA&A..35....1D}
}

@ARTICLE{2006AJ....132.2717M,
       author = {{Martin}, J.~C. and {Davidson}, Kris and {Koppelman}, M.~D.},
        title = "{The Chrysalis Opens? Photometry from the {\ensuremath{\eta}} Carinae Hubble Space Telescope Treasury Project, 2002-2006}",
      journal = {\aj},
         year = 2006,
        month = dec,
       volume = {132},
       number = {6},
        pages = {2717-2728},
          doi = {10.1086/508933},
archivePrefix = {arXiv},
       eprint = {astro-ph/0609295},
 primaryClass = {astro-ph},
       adsurl = {https://ui.adsabs.harvard.edu/abs/2006AJ....132.2717M}
}

@ARTICLE{2010ApJ...717L..22M,
       author = {{Mehner}, Andrea and {Davidson}, Kris and {Humphreys}, Roberta M. and {Martin}, John C. and {Ishibashi}, Kazunori and {Ferland}, Gary J. and {Walborn}, Nolan R.},
        title = "{A Sea Change in Eta Carinae}",
      journal = {\apjl},
         year = 2010,
        month = jul,
       volume = {717},
       number = {1},
        pages = {L22-L25},
          doi = {10.1088/2041-8205/717/1/L22},
archivePrefix = {arXiv},
       eprint = {1004.3529},
 primaryClass = {astro-ph.SR},
       adsurl = {https://ui.adsabs.harvard.edu/abs/2010ApJ...717L..22M}
}

@ARTICLE{2012Natur.486E...1D,
       author = {{Davidson}, Kris and {Humphreys}, Roberta M.},
        title = "{The Great Eruption of {\ensuremath{\eta}} Carinae}",
      journal = {\nat},
         year = 2012,
        month = jun,
       volume = {486},
       number = {7403},
        pages = {E1},
          doi = {10.1038/nature11166},
archivePrefix = {arXiv},
       eprint = {1205.2010},
 primaryClass = {astro-ph.SR},
       adsurl = {https://ui.adsabs.harvard.edu/abs/2012Natur.486E...1D}
}

@ARTICLE{2012ApJ...751...73M,
       author = {{Mehner}, Andrea and {Davidson}, Kris and {Humphreys}, Roberta M. and {Ishibashi}, Kazunori and {Martin}, John C. and {Ruiz}, Mar{\'\i}a Teresa and {Walter}, Frederick M.},
        title = "{Secular Changes in Eta Carinae's Wind 1998-2011}",
      journal = {\apj},
         year = 2012,
        month = may,
       volume = {751},
       number = {1},
          eid = {73},
        pages = {73},
          doi = {10.1088/0004-637X/751/1/73},
archivePrefix = {arXiv},
       eprint = {1112.4338},
 primaryClass = {astro-ph.SR},
       adsurl = {https://ui.adsabs.harvard.edu/abs/2012ApJ...751...73M}
}

@ARTICLE{2014A&A...564A..14M,
       author = {{Mehner}, Andrea and {Ishibashi}, Kazunori and {Whitelock}, Patricia and {Nagayama}, Takahiro and {Feast}, Michael and {van Wyk}, Francois and {de Wit}, Willem-Jan},
        title = "{Near-infrared evidence for a sudden temperature increase in Eta Carinae}",
      journal = {\aap},
         year = 2014,
        month = apr,
       volume = {564},
          eid = {A14},
        pages = {A14},
          doi = {10.1051/0004-6361/201322729},
archivePrefix = {arXiv},
       eprint = {1401.4999},
 primaryClass = {astro-ph.SR},
       adsurl = {https://ui.adsabs.harvard.edu/abs/2014A&A...564A..14M}
}

@INPROCEEDINGS{2012ASSL..384....1H,
       author = {{Humphreys}, Roberta M. and {Martin}, John C.},
        title = "{Eta Carinae: From 1600 to the Present}",
    booktitle = {Eta Carinae and the Supernova Impostors},
         year = 2012,
       editor = {{Davidson}, Kris and {Humphreys}, Roberta M.},
       series = {Astrophysics and Space Science Library},
       volume = {384},
        month = jan,
        pages = {1},
          doi = {10.1007/978-1-4614-2275-4_1},
       adsurl = {https://ui.adsabs.harvard.edu/abs/2012ASSL..384....1H}
}

@ARTICLE{1999SSRv...90..493I,
       author = {{Israelian}, G. and {de Groot}, M.},
        title = "{P Cygni: An Extraordinary Luminous Blue Variable}",
      journal = {\ssr},
         year = 1999,
        month = oct,
       volume = {90},
        pages = {493-522},
          doi = {10.1023/A:1005223314464},
archivePrefix = {arXiv},
       eprint = {astro-ph/9908309},
 primaryClass = {astro-ph},
       adsurl = {https://ui.adsabs.harvard.edu/abs/1999SSRv...90..493I}
}

@ARTICLE{1997A&A...326.1117N,
       author = {{Najarro}, F. and {Hillier}, D.~J. and {Stahl}, O.},
        title = "{A spectroscopic investigation of P Cygni. I. H and HeI lines.}",
      journal = {\aap},
         year = 1997,
        month = oct,
       volume = {326},
        pages = {1117-1134},
       adsurl = {https://ui.adsabs.harvard.edu/abs/1997A&A...326.1117N}
}

@ARTICLE{1988IrAJ...18..163D,
       author = {{de Groot}, Mart},
        title = "{The most luminous stars in the universe}",
      journal = {Irish Astronomical Journal},
         year = 1988,
        month = mar,
       volume = {18},
        pages = {163-170},
       adsurl = {https://ui.adsabs.harvard.edu/abs/1988IrAJ...18..163D}
}

@INPROCEEDINGS{2001ASPC..233..133N,
       author = {{Najarro}, F.},
        title = "{Spectroscopy of P Cygni}",
    booktitle = {P Cygni 2000: 400 Years of Progress},
         year = 2001,
       editor = {{de Groot}, M. and {Sterken}, C.},
       series = {Astronomical Society of the Pacific Conference Series},
       volume = {233},
        month = jun,
        pages = {133},
       adsurl = {https://ui.adsabs.harvard.edu/abs/2001ASPC..233..133N}
}

@ARTICLE{2020MNRAS.494..218R,
       author = {{Rivet}, J. -P. and {Siciak}, A. and {de Almeida}, E.~S.~G. and {Vakili}, F. and {Domiciano de Souza}, A. and {Fouch{\'e}}, M. and {Lai}, O. and {Vernet}, D. and {Kaiser}, R. and {Guerin}, W.},
        title = "{Intensity interferometry of P Cygni in the H {\ensuremath{\alpha}} emission line: towards distance calibration of LBV supergiant stars}",
      journal = {\mnras},
         year = 2020,
        month = may,
       volume = {494},
       number = {1},
        pages = {218-227},
          doi = {10.1093/mnras/staa588},
archivePrefix = {arXiv},
       eprint = {1910.08366},
 primaryClass = {astro-ph.IM},
       adsurl = {https://ui.adsabs.harvard.edu/abs/2020MNRAS.494..218R}
}

@ARTICLE{2007ApJ...666.1116S,
       author = {{Smith}, Nathan and {Li}, Weidong and {Foley}, Ryan J. and {Wheeler}, J. Craig and {Pooley}, David and {Chornock}, Ryan and {Filippenko}, Alexei V. and {Silverman}, Jeffrey M. and {Quimby}, Robert and {Bloom}, Joshua S. and {Hansen}, Charles},
        title = "{SN 2006gy: Discovery of the Most Luminous Supernova Ever Recorded, Powered by the Death of an Extremely Massive Star like {\ensuremath{\eta}} Carinae}",
      journal = {\apj},
         year = 2007,
        month = sep,
       volume = {666},
       number = {2},
        pages = {1116-1128},
          doi = {10.1086/519949},
archivePrefix = {arXiv},
       eprint = {astro-ph/0612617},
 primaryClass = {astro-ph},
       adsurl = {https://ui.adsabs.harvard.edu/abs/2007ApJ...666.1116S}
}

@ARTICLE{2010AJ....139.2269B,
       author = {{Balan}, Aurelian and {Tycner}, C. and {Zavala}, R.~T. and {Benson}, J.~A. and {Hutter}, D.~J. and {Templeton}, M.},
        title = "{The Spatially Resolved H{\ensuremath{\alpha}}-emitting Wind Structure of P Cygni}",
      journal = {\aj},
         year = 2010,
        month = jun,
       volume = {139},
       number = {6},
        pages = {2269-2278},
          doi = {10.1088/0004-6256/139/6/2269},
archivePrefix = {arXiv},
       eprint = {1004.0376},
 primaryClass = {astro-ph.SR},
       adsurl = {https://ui.adsabs.harvard.edu/abs/2010AJ....139.2269B}
}

@ARTICLE{2018NewA...65...29M,
       author = {{Michaelis}, Amir M. and {Kashi}, Amit and {Kochiashvili}, Nino},
        title = "{Periodicity in the light curve of P Cygni-Indication for a binary companion?}",
      journal = {\na},
         year = 2018,
        month = nov,
       volume = {65},
        pages = {29-34},
          doi = {10.1016/j.newast.2018.06.001},
archivePrefix = {arXiv},
       eprint = {1806.00769},
 primaryClass = {astro-ph.SR},
       adsurl = {https://ui.adsabs.harvard.edu/abs/2018NewA...65...29M}
}

@PROCEEDINGS{2012ASSL..384.....D,
        title = "{Eta Carinae and the Supernova Impostors}",
       editor = {{Davidson}, K. and {Humphreys}, R.},
    booktitle = {Eta Carinae and the Supernova Impostors},
         year = 2012,
       series = {Astrophysics and Space Science Library},
       volume = {384},
        month = jan,
          doi = {10.1007/978-1-4614-2275-4},
       adsurl = {https://ui.adsabs.harvard.edu/abs/2012ASSL..384.....D}

}

@ARTICLE{2011ApJ...738L...5P,
       author = {{Piro}, Anthony L.},
        title = "{g-mode Excitation During the Pre-explosive Simmering of Type Ia Supernovae}",
      journal = {\apjl},
         year = 2011,
        month = sep,
       volume = {738},
       number = {1},
          eid = {L5},
        pages = {L5},
          doi = {10.1088/2041-8205/738/1/L5},
archivePrefix = {arXiv},
       eprint = {1105.0936},
 primaryClass = {astro-ph.HE},
       adsurl = {https://ui.adsabs.harvard.edu/abs/2011ApJ...738L...5P}
}

@INPROCEEDINGS{1976IAUS...73...75P,
       author = {{Paczynski}, B.},
        title = "{Common Envelope Binaries}",
    booktitle = {Structure and Evolution of Close Binary Systems},
         year = 1976,
       editor = {{Eggleton}, Peter and {Mitton}, Simon and {Whelan}, John},
       series = {IAU Symposium},
       volume = {73},
        month = jan,
        pages = {75},
       adsurl = {https://ui.adsabs.harvard.edu/abs/1976IAUS...73...75P}
}

@ARTICLE{2018MNRAS.475.1198S,
       author = {{Soker}, Noam and {Gilkis}, Avishai},
        title = "{Explaining iPTF14hls as a common-envelope jets supernova}",
      journal = {\mnras},
         year = 2018,
        month = mar,
       volume = {475},
       number = {1},
        pages = {1198-1202},
          doi = {10.1093/mnras/stx3287},
archivePrefix = {arXiv},
       eprint = {1711.05180},
 primaryClass = {astro-ph.HE},
       adsurl = {https://ui.adsabs.harvard.edu/abs/2018MNRAS.475.1198S}
}

@ARTICLE{2004MNRAS.352.1213C,
       author = {{Chugai}, Nikolai N. and {Blinnikov}, Sergei I. and {Cumming}, Robert J. and {Lundqvist}, Peter and {Bragaglia}, Angela and {Filippenko}, Alexei V. and {Leonard}, Douglas C. and {Matheson}, Thomas and {Sollerman}, Jesper},
        title = "{The Type IIn supernova 1994W: evidence for the explosive ejection of a circumstellar envelope}",
      journal = {\mnras},
         year = 2004,
        month = aug,
       volume = {352},
       number = {4},
        pages = {1213-1231},
          doi = {10.1111/j.1365-2966.2004.08011.x},
archivePrefix = {arXiv},
       eprint = {astro-ph/0405369},
 primaryClass = {astro-ph},
       adsurl = {https://ui.adsabs.harvard.edu/abs/2004MNRAS.352.1213C}
}

@ARTICLE{2013Natur.494...65O,
       author = {{Ofek}, E.~O. and {Sullivan}, M. and {Cenko}, S.~B. and {Kasliwal}, M.~M. and {Gal-Yam}, A. and {Kulkarni}, S.~R. and {Arcavi}, I. and {Bildsten}, L. and {Bloom}, J.~S. and {Horesh}, A. and {Howell}, D.~A. and {Filippenko}, A.~V. and {Laher}, R. and {Murray}, D. and {Nakar}, E. and {Nugent}, P.~E. and {Silverman}, J.~M. and {Shaviv}, N.~J. and {Surace}, J. and {Yaron}, O.},
        title = "{An outburst from a massive star 40 days before a supernova explosion}",
      journal = {\nat},
         year = 2013,
        month = feb,
       volume = {494},
       number = {7435},
        pages = {65-67},
          doi = {10.1038/nature11877},
archivePrefix = {arXiv},
       eprint = {1302.2633},
 primaryClass = {astro-ph.HE},
       adsurl = {https://ui.adsabs.harvard.edu/abs/2013Natur.494...65O}
}

@ARTICLE{1997ARA&A..35..309F,
       author = {{Filippenko}, Alexei V.},
        title = "{Optical Spectra of Supernovae}",
      journal = {\araa},
         year = 1997,
        month = jan,
       volume = {35},
        pages = {309-355},
          doi = {10.1146/annurev.astro.35.1.309},
       adsurl = {https://ui.adsabs.harvard.edu/abs/1997ARA&A..35..309F}
}

@ARTICLE{2006MNRAS.372.1133G,
       author = {{Gomez}, H.~L. and {Dunne}, L. and {Eales}, S.~A. and {Edmunds}, M.~G.},
        title = "{Submillimetre emission from {\ensuremath{\eta}} Carinae}",
      journal = {\mnras},
         year = 2006,
        month = nov,
       volume = {372},
       number = {3},
        pages = {1133-1139},
          doi = {10.1111/j.1365-2966.2006.10921.x},
archivePrefix = {arXiv},
       eprint = {astro-ph/0608333},
 primaryClass = {astro-ph},
       adsurl = {https://ui.adsabs.harvard.edu/abs/2006MNRAS.372.1133G}
}

@ARTICLE{2019MNRAS.485..988O,
       author = {{Owocki}, Stanley P. and {Hirai}, Ryosuke and {Podsiadlowski}, Philipp and {Schneider}, Fabian R.~N.},
        title = "{Hydrodynamical simulations and similarity relations for eruptive mass-loss from massive stars}",
      journal = {\mnras},
         year = 2019,
        month = may,
       volume = {485},
       number = {1},
        pages = {988-1000},
          doi = {10.1093/mnras/stz461},
archivePrefix = {arXiv},
       eprint = {1902.06220},
 primaryClass = {astro-ph.SR},
       adsurl = {https://ui.adsabs.harvard.edu/abs/2019MNRAS.485..988O}
}

@ARTICLE{2020A&A...635A.127K,
       author = {{Kuriyama}, Naoto and {Shigeyama}, Toshikazu},
        title = "{Radiation hydrodynamical simulations of eruptive mass loss from progenitors of Type Ibn/IIn supernovae}",
      journal = {\aap},
         year = 2020,
        month = mar,
       volume = {635},
          eid = {A127},
        pages = {A127},
          doi = {10.1051/0004-6361/201937226},
archivePrefix = {arXiv},
       eprint = {1912.09738},
 primaryClass = {astro-ph.SR},
       adsurl = {https://ui.adsabs.harvard.edu/abs/2020A&A...635A.127K}
}

@ARTICLE{2021A&A...646A.118K,
       author = {{Kuriyama}, Naoto and {Shigeyama}, Toshikazu},
        title = "{Comparison between the first and second mass eruptions from progenitors of Type IIn supernovae}",
      journal = {\aap},
         year = 2021,
        month = feb,
       volume = {646},
          eid = {A118},
        pages = {A118},
          doi = {10.1051/0004-6361/202038637},
archivePrefix = {arXiv},
       eprint = {2006.06389},
 primaryClass = {astro-ph.SR},
       adsurl = {https://ui.adsabs.harvard.edu/abs/2021A&A...646A.118K}
}

@ARTICLE{2024ApJ...967...33C,
       author = {{Corso}, Nicholas J. and {Lai}, Dong},
        title = "{Mass Ejection Driven by Sudden Energy Deposition in Stellar Envelopes}",
      journal = {\apj},
         year = 2024,
        month = may,
       volume = {967},
       number = {1},
          eid = {33},
        pages = {33},
          doi = {10.3847/1538-4357/ad3e6c},
archivePrefix = {arXiv},
       eprint = {2401.09534},
 primaryClass = {astro-ph.SR},
       adsurl = {https://ui.adsabs.harvard.edu/abs/2024ApJ...967...33C}
}

@ARTICLE{2017MNRAS.470.1642F,
       author = {{Fuller}, Jim},
        title = "{Pre-supernova outbursts via wave heating in massive stars - I. Red supergiants}",
      journal = {\mnras},
         year = 2017,
        month = sep,
       volume = {470},
       number = {2},
        pages = {1642-1656},
          doi = {10.1093/mnras/stx1314},
archivePrefix = {arXiv},
       eprint = {1704.08696},
 primaryClass = {astro-ph.SR},
       adsurl = {https://ui.adsabs.harvard.edu/abs/2017MNRAS.470.1642F}
}

@ARTICLE{2018MNRAS.476.1853F,
       author = {{Fuller}, Jim and {Ro}, Stephen},
        title = "{Pre-supernova outbursts via wave heating in massive stars - II. Hydrogen-poor stars}",
      journal = {\mnras},
         year = 2018,
        month = may,
       volume = {476},
       number = {2},
        pages = {1853-1868},
          doi = {10.1093/mnras/sty369},
archivePrefix = {arXiv},
       eprint = {1710.04251},
 primaryClass = {astro-ph.SR},
       adsurl = {https://ui.adsabs.harvard.edu/abs/2018MNRAS.476.1853F}
}

@ARTICLE{2019ApJ...877...92O,
       author = {{Ouchi}, Ryoma and {Maeda}, Keiichi},
        title = "{Constraining Massive Star Activities in the Final Years through Properties of Supernovae and Their Progenitors}",
      journal = {\apj},
         year = 2019,
        month = jun,
       volume = {877},
       number = {2},
          eid = {92},
        pages = {92},
          doi = {10.3847/1538-4357/ab1a37},
archivePrefix = {arXiv},
       eprint = {1904.07878},
 primaryClass = {astro-ph.HE},
       adsurl = {https://ui.adsabs.harvard.edu/abs/2019ApJ...877...92O}
}

@ARTICLE{2018MNRAS.480.1466S,
       author = {{Smith}, Nathan and {Andrews}, Jennifer E. and {Rest}, Armin and {Bianco}, Federica B. and {Prieto}, Jose L. and {Matheson}, Tom and {James}, David J. and {Smith}, R. Chris and {Strampelli}, Giovanni Maria and {Zenteno}, A.},
        title = "{Light echoes from the plateau in Eta Carinae's Great Eruption reveal a two-stage shock-powered event}",
      journal = {\mnras},
         year = 2018,
        month = aug,
       volume = {480},
       number = {2},
        pages = {1466-1498},
          doi = {10.1093/mnras/sty1500},
archivePrefix = {arXiv},
       eprint = {1808.00992},
 primaryClass = {astro-ph.SR},
       adsurl = {https://ui.adsabs.harvard.edu/abs/2018MNRAS.480.1466S}
}

@ARTICLE{1967PhRvL..18..379B,
       author = {{Barkat}, Z. and {Rakavy}, G. and {Sack}, N.},
        title = "{Dynamics of Supernova Explosion Resulting from Pair Formation}",
      journal = {\prl},
         year = 1967,
        month = mar,
       volume = {18},
       number = {10},
        pages = {379-381},
          doi = {10.1103/PhysRevLett.18.379},
       adsurl = {https://ui.adsabs.harvard.edu/abs/1967PhRvL..18..379B}
}

@ARTICLE{2011ApJ...734..102K,
       author = {{Kasen}, Daniel and {Woosley}, S.~E. and {Heger}, Alexander},
        title = "{Pair Instability Supernovae: Light Curves, Spectra, and Shock Breakout}",
      journal = {\apj},
         year = 2011,
        month = jun,
       volume = {734},
       number = {2},
          eid = {102},
        pages = {102},
          doi = {10.1088/0004-637X/734/2/102},
archivePrefix = {arXiv},
       eprint = {1101.3336},
 primaryClass = {astro-ph.HE},
       adsurl = {https://ui.adsabs.harvard.edu/abs/2011ApJ...734..102K}
}

@ARTICLE{2016MNRAS.456.1320T,
       author = {{Takahashi}, Koh and {Yoshida}, Takashi and {Umeda}, Hideyuki and {Sumiyoshi}, Kohsuke and {Yamada}, Shoichi},
        title = "{Exact and approximate expressions of energy generation rates and their impact on the explosion properties of pair instability supernovae}",
      journal = {\mnras},
         year = 2016,
        month = feb,
       volume = {456},
       number = {2},
        pages = {1320-1331},
          doi = {10.1093/mnras/stv2649},
archivePrefix = {arXiv},
       eprint = {1511.03040},
 primaryClass = {astro-ph.SR},
       adsurl = {https://ui.adsabs.harvard.edu/abs/2016MNRAS.456.1320T}
}

@ARTICLE{2017ApJ...836..244W,
       author = {{Woosley}, S.~E.},
        title = "{Pulsational Pair-instability Supernovae}",
      journal = {\apj},
         year = 2017,
        month = feb,
       volume = {836},
       number = {2},
          eid = {244},
        pages = {244},
          doi = {10.3847/1538-4357/836/2/244},
archivePrefix = {arXiv},
       eprint = {1608.08939},
 primaryClass = {astro-ph.HE},
       adsurl = {https://ui.adsabs.harvard.edu/abs/2017ApJ...836..244W}
}

@ARTICLE{1995ApJS..101..181W,
       author = {{Woosley}, S.~E. and {Weaver}, Thomas A.},
        title = "{The Evolution and Explosion of Massive Stars. II. Explosive Hydrodynamics and Nucleosynthesis}",
      journal = {\apjs},
         year = 1995,
        month = nov,
       volume = {101},
        pages = {181},
          doi = {10.1086/192237},
       adsurl = {https://ui.adsabs.harvard.edu/abs/1995ApJS..101..181W}
}

@ARTICLE{1983A&A...126..207L,
       author = {{Langer}, N. and {Fricke}, K.~J. and {Sugimoto}, D.},
        title = "{Semiconvective diffusion and energy transport}",
      journal = {\aap},
         year = 1983,
        month = sep,
       volume = {126},
       number = {1},
        pages = {207},
       adsurl = {https://ui.adsabs.harvard.edu/abs/1983A&A...126..207L}
}

@ARTICLE{2006A&A...460..199Y,
       author = {{Yoon}, S. -C. and {Langer}, N. and {Norman}, C.},
        title = "{Single star progenitors of long gamma-ray bursts. I. Model grids and redshift dependent GRB rate}",
      journal = {\aap},
         year = 2006,
        month = dec,
       volume = {460},
       number = {1},
        pages = {199-208},
          doi = {10.1051/0004-6361:20065912},
archivePrefix = {arXiv},
       eprint = {astro-ph/0606637},
 primaryClass = {astro-ph},
       adsurl = {https://ui.adsabs.harvard.edu/abs/2006A&A...460..199Y}
}

@ARTICLE{2019A&A...625A.132S,
       author = {{Schootemeijer}, A. and {Langer}, N. and {Grin}, N.~J. and {Wang}, C.},
        title = "{Constraining mixing in massive stars in the Small Magellanic Cloud}",
      journal = {\aap},
         year = 2019,
        month = may,
       volume = {625},
          eid = {A132},
        pages = {A132},
          doi = {10.1051/0004-6361/201935046},
archivePrefix = {arXiv},
       eprint = {1903.10423},
 primaryClass = {astro-ph.SR},
       adsurl = {https://ui.adsabs.harvard.edu/abs/2019A&A...625A.132S}
}

@ARTICLE{2000A&A...360..952H,
       author = {{Herwig}, F.},
        title = "{The evolution of AGB stars with convective overshoot}",
      journal = {\aap},
         year = 2000,
        month = aug,
       volume = {360},
        pages = {952-968},
          doi = {10.48550/arXiv.astro-ph/0007139},
archivePrefix = {arXiv},
       eprint = {astro-ph/0007139},
 primaryClass = {astro-ph},
       adsurl = {https://ui.adsabs.harvard.edu/abs/2000A&A...360..952H}
}

@ARTICLE{2024ApJ...974..124M,
       author = {{Mukhija}, Bhawna and {Kashi}, Amit},
        title = "{Giant Eruptions in Massive Stars and their Effect on the Stellar Structure}",
      journal = {\apj},
         year = 2024,
        month = oct,
       volume = {974},
       number = {1},
          eid = {124},
        pages = {124},
          doi = {10.3847/1538-4357/ad7398},
archivePrefix = {arXiv},
       eprint = {2408.01718},
 primaryClass = {astro-ph.SR},
       adsurl = {https://ui.adsabs.harvard.edu/abs/2024ApJ...974..124M}
}

@ARTICLE{1973ApJ...181..429J,
       author = {{Joss}, P.~C. and {Salpeter}, E.~E. and {Ostriker}, J.~P.},
        title = "{On the ``Critical Luminosity'' in Stellar Interiors and Stellar Surface Boundary Conditions}",
      journal = {\apj},
         year = 1973,
        month = apr,
       volume = {181},
        pages = {429-438},
          doi = {10.1086/152060},
       adsurl = {https://ui.adsabs.harvard.edu/abs/1973ApJ...181..429J}
}

@ARTICLE{2011MNRAS.417.1466K,
       author = {{Kashi}, Amit and {Soker}, Noam},
        title = "{A circumbinary disc in the final stages of common envelope and the core-degenerate scenario for Type Ia supernovae}",
      journal = {\mnras},
         year = 2011,
        month = oct,
       volume = {417},
       number = {2},
        pages = {1466-1479},
          doi = {10.1111/j.1365-2966.2011.19361.x},
archivePrefix = {arXiv},
       eprint = {1105.5698},
 primaryClass = {astro-ph.SR},
       adsurl = {https://ui.adsabs.harvard.edu/abs/2011MNRAS.417.1466K}
}

@ARTICLE{2023ApJS..265...15J,
       author = {{Jermyn}, Adam S. and {Bauer}, Evan B. and {Schwab}, Josiah and {Farmer}, R. and {Ball}, Warrick H. and {Bellinger}, Earl P. and {Dotter}, Aaron and {Joyce}, Meridith and {Marchant}, Pablo and {Mombarg}, Joey S.~G. and {Wolf}, William M. and {Sunny Wong}, Tin Long and {Cinquegrana}, Giulia C. and {Farrell}, Eoin and {Smolec}, R. and {Thoul}, Anne and {Cantiello}, Matteo and {Herwig}, Falk and {Toloza}, Odette and {Bildsten}, Lars and {Townsend}, Richard H.~D. and {Timmes}, F.~X.},
        title = "{Modules for Experiments in Stellar Astrophysics (MESA): Time-dependent Convection, Energy Conservation, Automatic Differentiation, and Infrastructure}",
      journal = {\apjs},
         year = 2023,
        month = mar,
       volume = {265},
       number = {1},
          eid = {15},
        pages = {15},
          doi = {10.3847/1538-4365/acae8d},
archivePrefix = {arXiv},
       eprint = {2208.03651},
 primaryClass = {astro-ph.SR},
       adsurl = {https://ui.adsabs.harvard.edu/abs/2023ApJS..265...15J}
}

@ARTICLE{2025ApJ...994..159M,
       author = {{Martin}, John C. and {Humphreys}, Roberta M. and {Davidson}, Kris},
        title = "{On the Spatial Distribution of Luminous Blue Variables, B[e] Supergiants, and Wolf─Rayet Stars in the Large Magellanic Cloud}",
      journal = {\apj},
         year = 2025,
        month = dec,
       volume = {994},
       number = {2},
          eid = {159},
        pages = {159},
          doi = {10.3847/1538-4357/ae1021},
archivePrefix = {arXiv},
       eprint = {2508.17114},
 primaryClass = {astro-ph.SR},
       adsurl = {https://ui.adsabs.harvard.edu/abs/2025ApJ...994..159M}
}

@ARTICLE{2025Galax..13...29L,
       author = {{Lobel}, A. and {Gorlova}, N.},
        title = "{Multiplicity of Luminous Blue Variable Stars}",
      journal = {Galaxies},
         year = 2025,
        month = mar,
       volume = {13},
       number = {2},
          eid = {29},
        pages = {29},
          doi = {10.3390/galaxies13020029},
       adsurl = {https://ui.adsabs.harvard.edu/abs/2025Galax..13...29L}
}
\bibliographystyle{mnras}

\end{document}